\documentclass
[aps,superscriptaddress,nofootinbib,floatfix]{revtex4}%
\usepackage{amsfonts}
\usepackage{amsmath}
\usepackage{amssymb}
\usepackage{stackrel}
\usepackage{graphicx}
\usepackage[usenames,dvipsnames]{xcolor}%
\usepackage[utf8]{inputenc}
\usepackage[T1]{fontenc}
\usepackage[english]{babel}
\usepackage{tikz}
\providecommand{\U}[1]{\protect\rule{.1in}{.1in}}

\newcommand{\be}{\begin{equation}}
\newcommand{\ee}{\end{equation}}
\newcommand{\bea}{\begin{eqnarray}}
\newcommand{\eea}{\end{eqnarray}}

\newcommand{\bb}[1]{\left( #1 \right)}

\newcommand{\bbcro}[1]{\left[ #1 \right]}
\newcommand{\bbcror}[1]{\left. #1 \right]}
\newcommand{\bbcrol}[1]{\left[ #1 \right.}
\newcommand{\bbaco}[1]{\left\{ #1 \right\}}

\newcommand{\infinity}{\infty}

\renewcommand{\Re}{\textrm{Re}\,}
\renewcommand{\Im}{\textrm{Im}\,}

\begin{document}
\preprint{ }
\title{Phononic collective excitations in superfluid Fermi gases at nonzero temperatures}
\author{S. N. Klimin}
\affiliation{{TQC, Universiteit Antwerpen, Universiteitsplein 1, B-2610 Antwerpen, Belgium}}
\author{J. Tempere}
\affiliation{{TQC, Universiteit Antwerpen, Universiteitsplein 1, B-2610 Antwerpen, Belgium}}
\affiliation{{Lyman Laboratory of Physics, Harvard University, USA}}
\author{H. Kurkjian}
\affiliation{{TQC, Universiteit Antwerpen, Universiteitsplein 1, B-2610 Antwerpen, Belgium}}

\begin{abstract}
We study the phononic collective modes of the pairing field $\Delta$
and their corresponding signature in both the order-parameter and density response functions
for a superfluid Fermi gas at all temperatures below $T_c$ in the collisionless regime.
The spectra of collective modes are calculated within the Gaussian Pair Fluctuation approximation.
We deal with the coupling  of these modes to the fermionic continuum
of quasiparticle-quasihole excitations by performing a non-perturbative analytic continuation
of the pairing field propagator. At low temperature, we recover the known exponential temperature
dependence of the damping rate and velocity shift of the Anderson-Bogoliubov branch.
In the vicinity of $T_c$, we find analytically a weakly-damped collective mode
whose velocity vanishes with a critical exponent of $1/2$,
and whose quality factor diverges logarithmically with $T_c-T$, thereby clarifying an
existing debate in the literature (Andrianov \textit{et al.} Th. Math. Phys. \textbf{28}, 829,
Ohashi \textit{et al.} J. Phys. Jap.  \textbf{66}, 2437). A transition between these two
phononic branches is visible at intermediary temperatures, particularly in the BCS limit
where the phase-phase response function displays two maxima.
\end{abstract}
\maketitle

\section{Introduction \label{sec:intro}}

Collective excitations are sensitive probes of the microscopic physics of
many-body systems. In condensed cold gases, they can be experimentally
detected through response functions using for example Bragg spectroscopy. In
superfluid paired Fermi gases at zero temperature, they can be classified into several distinctive
branches: the Anderson-Bogoliubov (Goldstone) branch
\cite{Anderson,Ohashi2003,Combescot} which has a soundlike dispersion at low
momenta according to the Goldstone theorem; the \textit{pair-breaking
collective branch} in the pair-breaking continuum \cite{Popov1976,KurkjianPB} sometimes
referred to as Higgs (or Andrianov-Popov) branch; the Leggett branch in multiband systems
\cite{Leggett}. In this work, we focus on the phononic modes, which we characterize
in the collisionless regime at non zero temperature.

First predicted by Anderson \cite{Anderson} within the Random Phase
Approximation (RPA), phononic modes have been observed in a series of experiments
\cite{Bartenstein,Kinast,Altmeyer,Tey,Sidorenkov,Hoinka} over the last decade
and consequently became a subject of intensified theoretical investigation. At
zero temperature, the sound velocity \cite{Engelbrecht,Marini,Diener2008} and
later the full spectrum \cite{Diener2008,Ohashi2003,Combescot,Kurkjian2016} of the Anderson-Bogoliubov branch
were investigated theoretically in the Gaussian pair fluctuations (GPF)
approximation (equivalent to Anderson's RPA \cite{Kurkjian2017-2}). The obtained sound velocity agrees
with the first sound velocity predicted (in terms of the
gas compressibility) by dissipativeless quantum hydrodynamics \cite{Khalatnikov1949}.
Sophisticated low-energy effective theories were developed to go beyond hydrodynamics and
capture the first dispersive correction to the spectrum \cite{Manes,Rupak,Salasnich2008}, and disagree with GPF/RPA.
Finally, the finite lifetime of the phonons at $T=0$ was obtained by considering the Beliaev three-phonon
couplings \cite{Kurkjian2017}.

At $T\neq0$, the theory of collective excitations in Fermi gases
remains a very open field of investigation, stimulated by its relevance for
state-of-the-art experiments. On top of the bosonic couplings
\cite{Kurkjian2017,Escobedo}, which are known to cause the
temperature dependence of the energy of Bogoliubov quasiparticles
\cite{Beliaev} in a Bose gas, the collective excitations are
coupled, in a paired Fermi gas, to the fermionic quasiparticles or \textquotedblleft broken
pairs\textquotedblright\ \cite{KurkjianPB}. The GPF/RPA approximation is able
to describe the three-body process of absorption or emission of quasiparticles \cite{Ohashi2003,Zhang}
by the collective modes in the collisionless regime: the fermionic quasiparticles
are assumed to be non-interacting and thus (in the absence of impurities) to have
an infinite relaxation time. Pieri et al. \cite{Pieri2004}
showed that the exact resonance in the GPF response function, which characterizes
the collective mode at $T=0$, is replaced at nonzero temperature by a
broadened peak. Calculations of the phonon damping rate were performed
in the limit of low temperature \cite{Orbach1981,Kurkjian2017-2,Zou}, where it was found
to be exponentially small, with an activation energy strictly larger than the gap.
Close to the transition temperature, a phononic collective mode was also found \cite{Popov1976}
and its velocity was predicted to vanish as $(T_c-T)^{\alpha}$
with a critical exponent  $\alpha=1/2$ according to Ref.~\cite{Popov1976} and $\alpha=1/6$ according to Ref.~\cite{Takada1997}.

Collective modes have also been studied in the hydrodynamic limit where the relaxation
rate of the quasiparticles is much larger than the frequency of the collective mode, the
opposite of the collisionless limit considered here. In neutral Fermi systems,
this is done done using two-fluid hydrodynamics \cite{Martin1965} which predicts two phononic
branches, first and second sound. First sound coincides at $T=0$ with the Anderson-Bogolioubov mode found in collisionless theories
but this is no longer the case at $T\neq0$. First sound velocity remains nonzero at $T_c$, unlike the velocity of second sound which vanishes
near $T_c$ like the superfluid fraction. To describe the damping of these modes, hydrodynamic theories rely
on macroscopic dissipation coefficients \cite{Liu2018hydro} (the thermal conductivity and the various viscosities),
which have not been calculated theoretically, and are difficult to access experimentally.
In charged systems, the phononic nature of the Anderson-Bogoliubov mode is lost at low temperature
due to long-range Coulomb interactions that shift its energy toward the plasma energy \cite{Anderson}.
However, in dirty superconductors (which are far in the hydrodynamic
regime due to the presence of impurities), it was shown both experimentally
\cite{Goldman1973,Goldman1976} and theoretically \cite{Schon1975,Volkov1975,Volkov1979}
that a phononic collective mode, known as the Carlson-Goldman mode exists close to $T_c$.
The speed and damping rate of this collective mode were found to vanish at $T_c$.

In the present work, we compute the complex sound velocity of the phononic collective modes
within GPF in a \textit{self-consistent nonperturbative} way, which allows us to explore all temperatures from $0$ to $T_{c}$.
We show that the GPF effective action can be
rigorously expanded at low energy $\omega$ and wave number $q$ provided one
introduces a complex sound velocity $u$ and sets $\omega=uq$. The expansion
yields an explicit equation for $u$ that exhibits a branch cut for real $u$
due to the coupling between phonons and fermionic quasiparticles. Following
the procedure of Ref. \cite{KurkjianPB} for the pair-breaking branch, we solve
this equation after analytic continuation through the branch cut, and study
the solutions as functions of temperature and interaction strength. In
the limits $T\to0$ and $T\to T_c$, we perform this continuation entirely analytically.
For intermediate temperatures, we develop a numerical method to perform the analytic
continuation, which is based on the procedure of Nozi\`eres \cite{Nozieres}.

We find, in general, two complex roots to the dispersion equation. One root
describes the Anderson-Bogoliubov sound velocity in the
zero-temperature limit. Near the transition temperature, we find that there exists another phononic
collective mode whose complex velocity vanishes with a critical exponent of $1/2$ and
whose quality factor diverges logarithmically with $T_c-T$. This root appears in both the phase-phase and
density-density response functions as a resonance centered around $\omega/v_{\rm F}q\approx\Delta/T_c$,
which sharpens when approaching $T_c$. At intermediary temperature, the two phononic
branches coexist, and give a characteristic double-Lorentzian
shape (which is accentuated in the BCS regime) to the phase-phase response function.

Our results are in good agreement with the existing experimental data at low temperatures. In the vicinity of $T_c$,
where the order-parameter collective mode has not yet been observed, we explain how the
phase-phase response function could be measured by adapting to cold atoms the setup of Carlson and Goldman
based on a Josephson junction between a cold ($T\to0$) and a hot ($T\to T_c$) superfluid.

\section{Equation for the {complex} sound velocity \label{sec:equation}}

\subsection{Gaussian fluctuation action}

The present theoretical investigation of collective excitations in superfluid
Fermi gases is performed in the path-integral formalism. We consider ultracold
two-component Fermi gases with $s$-wave pairing, described
\cite{deMelo1993,Engelbrecht,Diener2008} by the action functional in Grassmann
variables $\left(  \bar{\psi}_{\sigma},\psi_{\sigma}\right)  $,%
\begin{equation}
S=\int_{0}^{\beta}d\tau\int d\mathbf{r}\left[  \sum_{\sigma=\uparrow
,\downarrow}\bar{\psi}_{\sigma}\left(  \frac{\partial}{\partial\tau}%
-\frac{\nabla_{\mathbf{r}}^{2}}{2m}-\mu\right)  \psi_{\sigma}+g\bar{\psi}_{\uparrow}%
\bar{\psi}_{\downarrow}\psi_{\downarrow}\psi_{\uparrow}\right]  , \label{S}%
\end{equation}
where $\beta=1/ T  $ is the inverse temperature (we set $\hbar=k_B=1$) and the
chemical potential $\mu$ fixes the total fermion density. The $s$-wave contact interactions are
characterized by the coupling constant $g<0$; the ultraviolet divergence of
the contact interaction model is removed by replacing $g$ by the $s$-wave
scattering length $a$ through the renormalization relation \cite{deMelo1993}:%
\begin{equation}
\frac{1}{g}=\frac{m}{4\pi a}-\int\frac{d^3k}{\left(  2\pi\right)  ^{3}%
}\frac{m}{k^{2}}. \label{g}%
\end{equation}

The further treatment is based on the effective bosonic pair field action
after the Hubbard-Stratonovich transformation with the pair field $\left[
\bar{\Psi},\Psi\right]  $ and the integration over the fermion fields, as in
\cite{deMelo1993,Engelbrecht,Diener2008}. This leads to the effective bosonic
action $S_{\mathrm{eff}}$ depending on the pair field only:%
\begin{equation}
S_{\mathrm{eff}}=-\operatorname{Tr}\ln\left[  -\mathbb{G}^{-1}\right]
-\int_{0}^{\beta}d\tau\int d\mathbf{r~}\frac{1}{g}\bar{\Psi}\left(
\mathbf{r},\tau\right)  \Psi\left(  \mathbf{r},\tau\right)  , \label{Seff1a}%
\end{equation}
where $\mathbb{G}^{-1}\left(  \mathbf{r},\tau\right)  $ is the inverse Nambu
tensor,%
\begin{equation}
\mathbb{G}^{-1}\left(  \mathbf{r},\tau\right)  =\left(
\begin{array}
[c]{cc}%
-\frac{\partial}{\partial\tau}+\frac{\nabla_{\mathbf{r}}^{2}}{2m}+\mu & \Psi\left(
\mathbf{r},\tau\right) \\
\bar{\Psi}\left(  \mathbf{r},\tau\right)  & -\frac{\partial}{\partial\tau
}-\frac{\nabla_{\mathbf{r}}^{2}}{2m}-\mu
\end{array}
\right)  . \label{Fa}%
\end{equation}

In the mean-field approximation, the pair field $\Psi\left(  \mathbf{r}%
,\tau\right)  $ is replaced by a uniform static order parameter $\Delta$,
solution of the mean-field gap equation
\begin{equation}
\int\frac{d^3k}{\left(  2\pi\right)  ^{3}}\frac{X\left(  E_{\mathbf{k}}\right)}{2E_{\mathbf{k}}%
}  +\frac{1}{g}=0. \label{Gapeq}%
\end{equation}
Here, $E_{\mathbf{k}}=\sqrt{\xi_{\mathbf{k}}^{2}+\Delta^{2}}$ is the energy of
the BCS quasiparticles, with $\xi_{\mathbf{k}}=k^{2}/2m-\mu$ the free fermion
energy. The temperature dependence comes in via the function%
\begin{equation}
X\left(  E_{\mathbf{k}}\right)  =\tanh\left(  \frac{\beta E_{\mathbf{k}}}%
{2}\right)  , \label{X}%
\end{equation}
related to the Fermi-Dirac occupation number $n(E_{\mathbf{k}})$ by
$X(E_{\mathbf{k}})=1-2n(E_{\mathbf{k}})$.
Finally, the mean-field critical temperature $T_{c}%
=1/\beta_{\mathrm{c}}$ is the temperature at which the order parameter $\Delta$ in Eq.
(\ref{Gapeq}) vanishes:
\begin{equation}
\int\frac{d^3k}{\left(  2\pi\right)  ^{3}}\frac{\tanh\left(  \frac{\beta_c \xi_{\mathbf{k}}}%
{2}\right)}{2\xi_{\mathbf{k}}%
}  +\frac{1}{g}=0. \label{Tc}%
\end{equation}

The Gaussian pair fluctuation approximation consists in expanding the action
(\ref{Seff1a}) to second order about the mean-field solution. The pair field
$\Psi$ is represented as a sum of the uniform and time-independent value
$\Delta$ and the fluctuation field $\varphi$:%
\begin{equation}
\Psi\left(  \mathbf{r},\tau\right)  =\Delta+\varphi\left(  \mathbf{r}%
,\tau\right)  ,\quad\bar{\Psi}\left(  \mathbf{r},\tau\right)  =\Delta
+\bar{\varphi}\left(  \mathbf{r},\tau\right)  \label{fluct}%
\end{equation}
and the fluctuations are taken into account up to second order. Next, the pair
field action is rewritten in Fourier space with variables $\left(
\mathbf{q},i\Omega_{n}\right)  $ where $\Omega_{n}=2\pi n/\beta$ is the
bosonic Matsubara frequency. This gives us the quadratic fluctuation action in
matrix form:%
\begin{equation}
S^{\left(  \mathrm{quad}\right)  }=\frac{1}{2}\sum_{\mathbf{q},n}\left(
\begin{array}
[c]{cc}%
\bar{\varphi}_{\mathbf{q},n} & \varphi_{-\mathbf{q},-n}%
\end{array}
\right)  \mathbb{M}\left(  \mathbf{q},i\Omega_{n}\right)  \left(
\begin{array}
[c]{c}%
\varphi_{\mathbf{q},n}\\
\bar{\varphi}_{-\mathbf{q},-n}%
\end{array}
\right)  , \label{Squad}%
\end{equation}
with the inverse fluctuation propagator $\mathbb{M}\left(  \mathbf{q}%
,i\Omega_{n}\right)  $. The collective modes of the system are the eigenmodes
of the quadratic action (\ref{Squad}). The explicit form of the matrix
elements of $\mathbb{M}$ with the coupling constant renormalized according to (\ref{g})
reads:
\begin{align}
M_{1,1}\left(  \mathbf{q},i\Omega_{n}\right)   &  =M_{2,2}\left(
-\mathbf{q},-i\Omega_{n}\right) \nonumber\\
&  {=\int\frac{d^3 k}{\left(  2\pi\right)  ^{3}}\left[  \frac
{X(E_{\mathbf{k}})}{2E_{\mathbf{k}}}+\frac{X\left(  E_{\mathbf{k}}\right)
}{4E_{\mathbf{k}}E_{\mathbf{k}+\mathbf{q}}}\right.  }\nonumber\\
&  \times\left(  \frac{\left(  \xi_{\mathbf{k}}+E_{\mathbf{k}}\right)  \left(
E_{\mathbf{k}+\mathbf{q}}+\xi_{\mathbf{k}+\mathbf{q}}\right)  }{i\Omega
_{n}-E_{\mathbf{k}}-E_{\mathbf{k}+\mathbf{q}}}-\frac{\left(  \xi_{\mathbf{k}%
}-E_{\mathbf{k}}\right)  \left(  \xi_{\mathbf{k}+\mathbf{q}}-E_{\mathbf{k}%
+\mathbf{q}}\right)  }{i\Omega_{n}+E_{\mathbf{k}}+E_{\mathbf{k}+\mathbf{q}}%
}\right.  \nonumber\\
&  \left.  \left.  -\frac{\left(  \xi_{\mathbf{k}}+E_{\mathbf{k}}\right)
\left(  \xi_{\mathbf{k}+\mathbf{q}}-E_{\mathbf{k}+\mathbf{q}}\right)
}{i\Omega_{n}-E_{\mathbf{k}}+E_{\mathbf{k}+\mathbf{q}}}+\frac{\left(
\xi_{\mathbf{k}}-E_{\mathbf{k}}\right)  \left(  \xi_{\mathbf{k}+\mathbf{q}%
}+E_{\mathbf{k}+\mathbf{q}}\right)  }{i\Omega_{n}+E_{\mathbf{k}}%
-E_{\mathbf{k}+\mathbf{q}}}\right)  \right]  , \label{M11}%
\end{align}
and%
\begin{align}
M_{1,2}\left(  \mathbf{q},i\Omega_{n}\right)   &  =M_{2,1}\left(
-\mathbf{q},-i\Omega_{n}\right) \nonumber\\
&  =-\Delta^{2}\int\frac{d^3 k}{\left(  2\pi\right)  ^{3}}\frac{X\left(
E_{\mathbf{k}}\right)  }{4E_{\mathbf{k}}E_{\mathbf{k}+\mathbf{q}}}\nonumber\\
&  \times\left(  \frac{1}{i\Omega_{n}-E_{\mathbf{k}}-E_{\mathbf{k}+\mathbf{q}%
}}-\frac{1}{i\Omega_{n}+E_{\mathbf{k}}+E_{\mathbf{k}+\mathbf{q}}}\right.
\nonumber\\
&  \left.  -\frac{1}{i\Omega_{n}-E_{\mathbf{k}}+E_{\mathbf{k}+\mathbf{q}}%
}+\frac{1}{i\Omega_{n}+E_{\mathbf{k}}-E_{\mathbf{k}+\mathbf{q}}}\right)  .
\label{M12}%
\end{align}
Note that the \textquotedblleft quasiparticle-quasihole\textquotedblright%
\ parts of the matrix coefficients with denominator $i\Omega_{n}\pm\left(
E_{\mathbf{k}}-E_{\mathbf{k}+\mathbf{q}}\right)  $ vanish at $T=0$ [where
$X(E_{\mathbf{k}})=1$] as can be seen by the change of variable $\mathbf{k}%
\leftrightarrow-\mathbf{k}-\mathbf{q}$, and at $T=T_{c}$ since in this case
$E_{\mathbf{k}}=\xi_{\mathbf{k}}$.

\subsection{Spectrum of the collective modes}

The complex energies $z_{\mathbf{q}}$ of the collective excitations can be
determined as the complex poles of the fluctuation propagator $z\mapsto
\mathbb{M}^{-1}(q,z)$, or, equivalently, as the complex roots of the
determinant of $\mathbb{M}$:%
\begin{equation}
\det\mathbb{M}\left(  q,z_{\mathbf{q}}\right)  =0. \label{detMeq0}%
\end{equation}
One usually separates in $z_{\mathbf{q}}$ the real part and imaginary part:
\begin{equation}
z_{\mathbf{q}}=\omega_{\mathbf{q}}-i\Gamma_{\mathbf{q}}/2, \label{zq}%
\end{equation}
where $\omega_{\mathbf{q}}$ is the mode frequency and $\Gamma_{\mathbf{q}}$
its damping rate.

The straightforward analytic continuation of the matrix coefficients
(\ref{M11}) and (\ref{M12}) by the replacement $i\Omega_{n}\rightarrow z$ has
a branch cut along the \textit{whole} real axis (unlike in the $T=0$ case
\cite{KurkjianPB} where the branch cut begins at $2\Delta$) due to the
denominator $z\pm\left(  E_{\mathbf{k}}-E_{\mathbf{k}+\mathbf{q}}\right)  $.
The roots of Eq.~\eqref{detMeq0}, even the low-energy ones, can then only be
found when the determinant is analytically continued through the branch cut
following the method proposed by Nozi\`{e}res \cite{Nozieres,KurkjianPB}. The
aim of this paper is to perform this analytic continuation and track the
low-energy solutions of (\ref{detMeq0}) in the complex $z$-plane as functions
of interaction strength and temperature.

\subsection{Equation of state}
\label{sec:eos}
In dimensionless form, the Gaussian fluctuation matrix $\mathbb{M}$,
and hence the collective mode energy $z_{\bf q}$, depend on two reduced parameters: $\Delta/T$
 and $\Delta/\mu$, which both depend on temperature.
 One may want to replace these parameters by more usual quantities such
as $T/T_c$, and the interaction strength,
measured by the product $k_{ F}a$ of the scattering length $a$ and Fermi wavevector $k_{ F}$.
This is done in three steps. First, one uses the number equation to express
$\Delta/\epsilon_{ F}$ ($\epsilon_{ F}$ is the Fermi energy) as a function of $\Delta/T$ and $\Delta/\mu$.
The crudest approximation, which can be used only for a qualitative explanation of collective excitations
is the mean-field number equation
\begin{equation}
n\equiv\frac{k_{ F}^3}{3\pi^{2}}{=\int\frac{d^3{ k}}{(2\pi)^3}\left(  1-\frac
{\xi_{\mathbf{k}}}{E_{\mathbf{k}}}X(E_{\mathbf{k}})\right)  .} \label{eos}%
\end{equation}
where $n$ is the average density of the gas. Second, one relates $k_{ F}a$ to
$\Delta/T$, $\Delta/\mu$ and $\Delta/\epsilon_{\rm F}$ by combining Eqs.~\eqref{g} and \eqref{Gapeq}:
\be
\frac{m}{4\pi a}=\int\frac{d^3{ k}}{(2\pi)^3}\bb{\frac{m}{k^2}-\frac{X\left(  E_{\mathbf{k}}\right)}{2E_{\mathbf{k}}}}.
\label{gap2}
\ee
With these two equations, one can change the parametrization of $z_{\bf q}$ from ($\Delta/T$,$\Delta/\mu$)
to $(k_{\rm F}a,\Delta/\epsilon_{\rm F})$, or equivalently to $(k_{\rm F}a,T/\epsilon_{\rm F})$ using $T/\epsilon_{\rm F}=\Delta/\epsilon_{\rm F}\times T/\Delta$.
Third, there remains to express $T_{\rm c}/\epsilon_{\rm F}$ as a function of $k_{\rm F}a$ using Eqs.~\eqref{eos}
and \eqref{gap2} specified at $T=T_{c}$, that is for $\Delta=0$: Eq.~\eqref{eos} yields $T_{\rm c}/\epsilon_{\rm F}$ as a function of $\mu(T_{\rm c})/T_{\rm c}$
and Eq.~\eqref{gap2} relates this last parameter to $k_{\rm F}a$.

In this process, the mean-field number equation \eqref{eos} can be
replaced by a more accurate one, such as the number
the number equation obtained via renormalization group theory \cite{Boettcher},
the one obtained from Monte Carlo calculations \cite{Astr,Bulgac2006}, or the one extracted from
experimental data \cite{Salomon2010,Zwierlein2012,Hoinka}.
Here we will use more particularly an equation of state which incorporates the
Gaussian fluctuations of the order-parameter [Eq.~\eqref{Squad}] to the number
equation, as proposed in Refs.~\cite{Nozieres1985,Diener2008,HLD,HLD1}.
This allows us to avoid the aberrant mean-field prediction of a diverging
$T_c$ in the BEC regime \cite{Nozieres1985}.
A major issue of these equations of state accounting for Gaussian fluctuations
is that they lead to artifacts when using the mean-field gap equation
\eqref{gap2} near $T_c$ (they loose the $\Delta=O(\sqrt{T_c-T})$ critical behavior
known from the theory of Ginzburg-Landau and predict an aberrant first order phase transition).
As explained in Appendix \ref{app:eos}, we solve this issue by rescaling the temperature at which the
gap equation is used by the ratio of the mean-field and corrected critical temperature
(which is a refinement of the idea of Refs.~\cite{Taylor2008,Taylor}). In this way, the zero temperature equation-of-state
coincides with the ``GPF'' scheme of Ref.~\cite{HLD}, the critical temperature is
the one computed by Nozi\`eres -- Schmitt-Rink \cite{Nozieres1985}, and the
critical behavior $\Delta=O(\sqrt{T_c-T})$ (which is crucial for our study of the collective modes)
is preserved.

Using an improved equation of state does not qualitatively
change our results on the collective modes (it is a mere rescaling
of the dependence on $k_{\rm F}a$ and $T/T_c$) but makes them more quantitative.
This strategy is used in our numeric results for collective excitations,
particularly in Secs.~\ref{sec:LowT},~\ref{sec:analcont}
for the spectra of collective modes
and in Sec.~\ref{sec:comparison} to compare our results to measurements of
the sound velocity.

\section{Long-wavelength expansion \label{sec:expansion}}

\subsection{Expansion of the $\mathbb{M}$ matrix for phononic energies}

In the present treatment, we focus on obtaining an analytic expression of the
velocity of the phononic modes in the long-wavelength limit ($q\rightarrow0$). Their eigenenergy
is expected to behave as $z_{\mathbf{q}}\sim u_{\mathrm{s}}q$,
with the complex sound velocity $u_{\mathrm{s}}$. The sound velocity
was calculated at $T=0$ \cite{Marini,Salasnich,Diener2008}
where the quasiparticle-quasihole branch cut vanishes such that $u_{\mathrm{s}}$ is real
and the long-wavelength expansion of the matrix elements $M_{j,k}\left(
\mathbf{q},z\right)  $, $j,k=1,2$ presents no difficulty, i. e., the
two-dimensional expansion in powers of $q$ and $z$ can be done successively.
Predictions of the limiting behavior at the transition temperature ($T\to T_c$)
are also available \cite{Popov1976,Takada1997} at weak coupling, and will be discussed
in section \ref{sec:Tc}.

For $0<T<T_{c}$, the point $(q=0,z=0)$ is a branch point of {$\det\mathbb{M}$}
and different limiting values when $(q,z)\rightarrow(0,0)$ can be obtained
depending on the path followed in the $(q,z)$ hyperplane. Therefore, there
exists no Taylor expansion valid everywhere in a vicinity of the point
$(q=0,z=0)$ \cite{Engelbrecht}. An expansion can be obtained nonetheless
assuming that $q$ and $z$ are small yet proportional to
each other. Consequently, we set $z\equiv uq$, where $u$ is a complex number
independent of $q$. An analogous trick was performed in Ref. \cite{Kosztin}.

In the $q\rightarrow0$ limit, it is more tractable to express the matrix
elements \eqref{M11} and \eqref{M12} in the modulus-phase basis,
\begin{equation}
\tilde{\mathbb{{M}}}\left(  \mathbf{q},z\right)  =\left(
\begin{array}
[c]{cc}%
{M}_{--}\left(  \mathbf{q},z\right)   & {M}_{+-}\left(
\mathbf{q},z\right)  \\
{M}_{-+}\left(  \mathbf{q},z\right)   & {M}_{++}\left(
\mathbf{q},z\right)
\end{array}
\right),  \label{mmatr}%
\end{equation}
where the new matrix elements are obtained by the unitary transformation \cite{Engelbrecht}:%
\begin{align}
{M}_{++}\left(  \mathbf{q},z\right)   &  =\frac{M_{1,1}\left(
\mathbf{q},z\right)  +M_{1,1}\left(  \mathbf{q},-z\right)  }{2}-M_{1,2}\left(
\mathbf{q},z\right)  ,\\
{M}_{--}\left(  \mathbf{q},z\right)   &  =\frac{M_{1,1}\left(
\mathbf{q},z\right)  +M_{1,1}\left(  \mathbf{q},-z\right)  }{2}+M_{1,2}\left(
\mathbf{q},z\right)  ,\\
{M}_{+-}\left(  \mathbf{q},z\right)   &  =\frac{M_{1,1}\left(
\mathbf{q},z\right)  -M_{1,1}\left(  \mathbf{q},-z\right)  }{2}=-{M}_{-+}\left(  \mathbf{q},z\right).
\end{align}
The diagonal matrix elements ${M}_{++}\left(  \mathbf{q},z\right)  $
and ${M}_{--}\left(  \mathbf{q},z\right)  $ correspond to the phase
and modulus fluctuations, respectively. The nondiagonal matrix elements describe
mixing of modulus and phase fluctuations. The series expansion in powers of
$q$ in this basis gives:%
\bea
{M}_{++}(\mathbf{q},uq)\!&=&\!\frac{q^2}{2m\Delta}\frac{m_{++}(u)}{\Delta} +O(q^4),\\
{M}_{--}(\mathbf{q},uq) \!&=&\frac{m_{--}(u)}{\Delta}+O(q^2),\\
{M}_{+-}(\mathbf{q},uq)\!&=&\!\frac{uq }{\Delta} \frac{m_{+-}(u)}{\Delta} +O(q^3),
\eea
with coefficients (dimensionless except for the Jacobian $d^3k$):
\bea
m_{++}(u)&=&\int\frac{\Delta^2d^3{ k}}{(2\pi)^3} \bbcrol{ \frac{e_c(v_{\bf k})}{6}\bb{\frac{ X(E_{\bf k})}{E_{\bf k}^3}-\frac{ X'(E_{\bf k})}{E_{\bf k}^2}}
 -\frac{e_c(u)}{2}\frac{ X(E_{\bf k})}{E_{\bf k}^3}} \notag\\
&&\bbcror{\qquad\qquad\qquad\qquad\quad+  e_c(u)\frac{\Delta^2}{2{E}_{\bf k}^2}\frac{  X'(E_{\bf k})e_c(v_{\bf k})\cos^2\theta}{({E}_{\bf k}^2e_c({u})-{\xi}_{\bf k}^2e_c(v_{\bf k})\cos^2\theta)}}, \label{mpp}\\
m_{--}(u) &=& \int\frac{\Delta^3d^3{ k}}{(2\pi)^3} \bbcro{ \frac{X(E_{\bf k})}{2{E}_{\bf k}^3}
+  \frac{{\xi}_{\bf k}^2}{2{E}_{\bf k}^2}\frac{ X'(E_{\bf k})e_c(v_{\bf k})\cos^2\theta}{({E}_{\bf k}^2e_c({u})-{\xi}_{\bf k}^2e_c(v_{\bf k})\cos^2\theta)}}, \label{mmm}\\
m_{+-}(u) &=&\int\frac{\Delta^2d^3{ k}}{(2\pi)^3} \bbcro{- \frac{{\xi}_{\bf k}X(E_{\bf k})}{4{E}_{\bf k}^3}
+\frac{\Delta{\xi}_{\bf k}}{4{E}_{\bf k}^2}\frac{ \Delta X'(E_{\bf k})e_c(v_{\bf k})\cos^2\theta}{({E}_{\bf k}^2e_c({u})-{\xi}_{\bf k}^2e_c(v_{\bf k})\cos^2\theta)}}. \label{mpm}
\eea
Here $v_{\bf k}=k/m$ is the phase velocity associated to wave vector $k$, and $e_c(v)=mv^2/2$ is the kinetic energy associated to velocity $v$.

\subsection{Reduced dispersion equation}

Substituting the series expansions of the matrix elements into the determinant
of $\mathbb{\tilde{M}}$, we get%
\begin{equation}
\det\mathbb{\tilde{M}}\left(  \mathbf{q},z=uq\right)  ={W\left(  u\right)}
\frac{q^{2}}{2m\Delta^3}+O\left(  q^{4}\right),
\end{equation}
where the function $W\left(  u\right)  $ is given by:
\begin{equation}
W\left(  u\right)  =  m_{++}(u)m_{--}(u)-\frac{2mu^2}{\Delta}m_{+-}^2(u), \label{fdet}%
\end{equation}
Let $u_{\mathrm{s}}$ be a generic solution of the low-$q$ dispersion equation:
\begin{equation}
W\left(  u_{\mathrm{s}}\right)  =0. \label{Wu}%
\end{equation}
The real part of $u_{\mathrm{s}}$ is readily interpreted as a sound
velocity
\begin{equation}
c_{\mathrm{s}}\equiv\text{$\operatorname{Re}$}\left(  u_{\mathrm{s}}\right)
=\lim_{q\rightarrow0}\frac{\omega_{q}}{q}%
\end{equation}
and the imaginary part
\begin{equation}
\kappa_{\mathrm{s}}\equiv\operatorname{Im}\left(  u_{\mathrm{s}}\right)
=\lim_{q\rightarrow0}\frac{\Gamma_{q}}{2q}%
\end{equation}
gives access to the long-wavelength limit of an inverse quality factor
$\Gamma_{q}/\omega_{q}$:%
\begin{equation}
\frac{2\kappa_{\mathrm{s}}}{c_{\mathrm{s}}}=\lim_{q\rightarrow0}\frac
{\Gamma_{q}}{\omega_{q}}. \label{limits}%
\end{equation}

As such, the reduced dispersion equation \eqref{Wu} has no root:
none on the real axis ($u\in\mathbb{R}$) which is entirely spanned by the branch cut
caused by the resonant denominator in Eqs.~(\ref{mpp}--\ref{mpm}), and none either in
the lower complex plane ($\textrm{Im}\,u<0$), otherwise there would also exist an unstable solution
in the upper plane (since $W(u)=0\implies W(-u)=0$). Two distinct strategies
can be adopted to overcome this apparent paradox.

$(i)$ One can limit the study to the vicinity of the real
axis setting $u=c+i0^+$ with $c\in\mathbb{R}$, and study the various responses of the system as a function of $c$.
Although the response functions (defined in the next subsection) have no pole, they may exhibit resonance peaks whose position and width
may be fitted to extract the real and imaginary parts of a phenomenological speed of sound.
This corresponds to an experiment where the response of the gas is
recorded at fixed (and low) $q$ as a function of $\omega$, using for example Bragg spectroscopy \cite{Hoinka}.
The disadvantage of this strategy is that it relies on a delicate choice of a
fitting function  \cite{Takada1997} for $1/W(c+i0^+)$, in particular in the case (that we will encounter) where the function
has more than one peak.

$(ii)$ One can instead look for true solutions of the dispersion
equation \eqref{Wu} in the analytic continuation through the branch cut. Knowledge of the poles of $1/W(u)$
in the complex plane makes it easy to devise  an analytic approximation for the response functions.
It also allows for a clear definition of the speed of sound,
and therefore for a rigorous study of its temperature dependence,
and in particular of its critical exponent near $T_c$.

\subsection{Response functions}

The response functions of the pair field in the GPF approximation are the
coefficients of the propagator {$\mathbb{\tilde{M}}^{-1}$} evaluated on the
real axis $z=\omega+i0^{+}$ \cite{Castin2001} (hence without analytic continuation through the
branch cut). In the low-$q$ limit, $\tilde{\mathbb{{M}}}^{-1}\left(
\mathbf{q},z\right)  $ is given by:%
\begin{equation}
\frac{{q}^{2}}{2m}\mathbb{\tilde{M}}^{-1}\left(  \mathbf{q},uq\right)
=\frac{\Delta}{W\left(  u\right)  }\left(
\begin{array}
[c]{cc}%
m_{++}(u)\frac{q^{2}}{2m} & -m_{+-}(u)uq\\
-m_{+-}(u)uq & \Delta m_{--}(u)
\end{array}
\right)  +O\left(  q^{4}\right)  .
\end{equation}
The largest response is thus in the phase-phase propagator $[\tilde
{\mathbb{{M}}}^{-1}]_{2,2}$. We define
\begin{equation}
\chi(c)\equiv\lim_{q\rightarrow0}\frac{1}{\pi}\operatorname{Im}\left\{
\frac{q^{2}}{2m\Delta^{2}}[\mathbb{\tilde{M}}^{-1}]_{2,2}(\mathbf{q}%
,(c+i0^{+})q)\right\}  =\frac{1}{\pi}\operatorname{Im}\frac{m_{--}\left(
c+i0^{+}\right)  }{W\left(  c+i0^{+}\right)  },\label{hilw}%
\end{equation}
the phase-phase response as a function of the velocity $c=\omega
/q\in\mathbb{R}$.

To account for density excitations, one should supplement the quadratic action
\eqref{Squad} by an auxiliary action containing the exciting
density fields, which we do in Appendix \ref{structfactor}. The result is the following expression of the
retarded density-density Green's function\footnote{We omit the $+i0^+$ (needed to avoid the branch cut) after $\omega$ on the right-hand side.}
\begin{align}
{G}_{\rho}^{R}\left(  \mathbf{q},\omega+i0^+\right)   &  ={M}%
_{\rho\rho}\left(  \mathbf{q},\omega\right) \nonumber\\
&  -\left\{  \left[  {M}_{-\rho}\left(  \mathbf{q},\omega\right)
\right]  ^{2}\frac{{M}_{++}\left(  \mathbf{q},\omega\right)
}{\det\mathbb{\tilde{M}}\left(  \mathbf{q},\omega\right)  }+\left[
{M}_{+\rho}\left(  \mathbf{q},\omega\right)  \right]  ^{2}%
\frac{{M}_{--}\left(  \mathbf{q},\omega\right)  }{\det
\mathbb{\tilde{M}}\left(  \mathbf{q},\omega\right)  }\right. \nonumber\\
&  \left.  -2{M}_{-\rho}\left(  \mathbf{q},\omega\right)
{M}_{+\rho}\left(  \mathbf{q},i\Omega_{m}\right)  \frac{{M}_{+-}\left(
\mathbf{q},\omega\right)  }{\det\mathbb{\tilde{M}}\left(  \mathbf{q}%
,\omega\right)  }\right\}  . \label{Gr}%
\end{align}
which is in agreement with Eq.~(20) of Ref.~\cite{Castin2001} (taking the density-density element of the response function matrix).
The density-pairing field and density-density elements of the fluctuation matrix, $M_{\pm\rho}$ and $M_{\rho\rho}$, (and their low-$q$ expansion) are given explicitly in Appendix \ref{structfactor}.
We then define the low-$q$ density-density response function as
\begin{equation}
\chi_{\rho}\left(  c\right)  =-\frac{1}{\pi}\lim_{q\rightarrow0}\left[
\operatorname{Im}G_{\rho}^{R}\left(  \mathbf{q},\left(  c+i0^{+}\right)
q\right)  \right]  . \label{lwlim}%
\end{equation}
It is related to the long wavelength density-density response function
by $\lim_{q\rightarrow0}S(\textbf{q},cq)=\chi_{\rho}\left(  c\right)/(1-e^{-\beta cq})$.
$\chi_{\rho}$ is composed of two terms which have a distinct physical origin
\begin{equation}
\chi_{\rho}\left(  c\right)  =\chi_{\rho}^{\left(  1\right)  }\left(
c\right)  +\chi_{\rho}^{\left(  2\right)  }\left(  c\right)  ,\label{subdiv}%
\end{equation}
The first term,%
\begin{equation}
\chi_{\rho}^{\left(  1\right)  }\left(  c\right)  =-\frac{1}{\pi}%
\lim_{q\rightarrow0}\left[  \operatorname{Im}{M}_{\rho\rho}\left(
\mathbf{q},\left(  c+i0^{+}\right)  q\right)  \right]  \label{first}%
\end{equation}
does not disappear at $T_c$; above it, it describes the known density response
of free fermions. The second contribution $\chi_{\rho}^{\left(  2\right)  }$ gathers
the terms between curly brackets in \eqref{Gr}, which have the determinant of $\tilde{\mathbb{M}}$ (the pairing field
fluctuation matrix) in the denominator. It describes the contribution of the pairing field to the density response
and it is specific to the superfluid phase. Due to the $\det\mathbb{\tilde{M}}$ in the denominator of this term,
it has the same poles and thus the same collective modes as the pair field response function.

\section{Low-temperature behavior \label{sec:LowT}}

\begin{figure}[tbh]%
\centering
\includegraphics[
width=3.5in
]%
{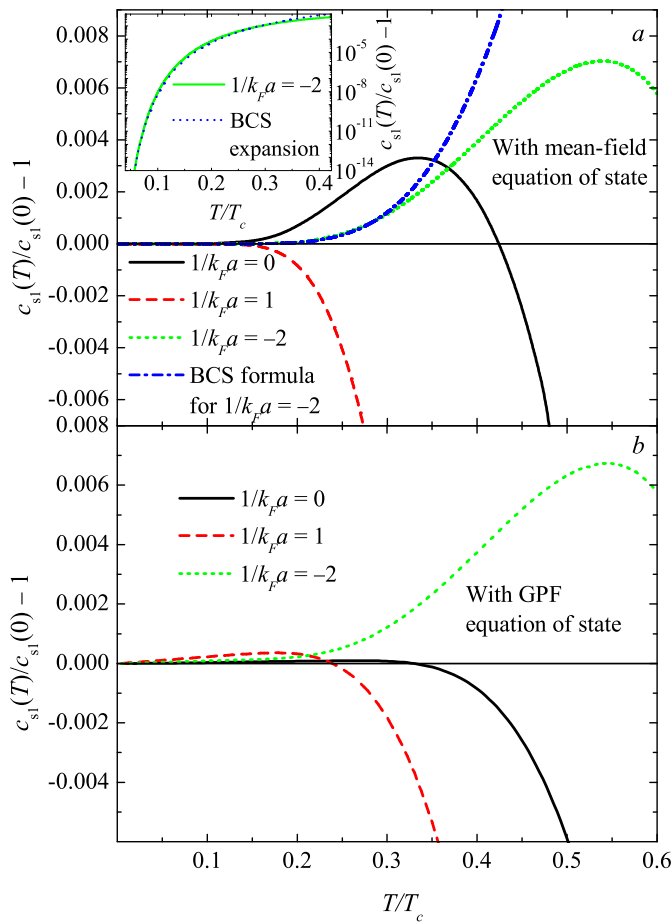}%
\caption{(Color online) (\textit{a}) Relative correction to the sound velocity
$c_{\mathrm{s}1}$ at low temperatures $T\ll \Delta$, calculated using the mean-field equation of state at unitarity
($1/|a|=0$ solid curve), in the BEC regime ($1/k_F a=1$, dashed red curve)
and in the BCS regime ($1/k_F a=-2$, green dashed curve). This last curve is compared
to the low-temperature exponential formula of \cite{Orbach1981}, (blue dashed-dotted curve),
also in the inset in logarithmic scale.
(\textit{b}) These corrections calculated using the GPF equation of state.}%
\label{fig:svelsLT}%
\end{figure}

We briefly study the behavior of the speed of sound at zero and low temperature, which are overall
well established results.
At $T=0$, one has $X(E_{k})=1$ and $X'(E_{k})=0$ such that the coefficients
(\ref{mpp}--\ref{mpm}) of the $\left(  q,z\right)  $ expansion depend trivially on
$u$ (as expected since the singular \textquotedblleft
quasiparticle-quasihole\textquotedblright\ terms vanish). The dispersion equation
(\ref{Wu}) in this case has one real root $u_{\mathrm{s,0}}(T=0)=c_{\mathrm{s,0}%
}(T=0)$ which satisfies the hydrodynamic formula $mc_{\mathrm{s,0}}^{2}%
=nd\mu/dn$ \cite{Martin1965,Taylor} and can thus be unambiguously identified as the first
sound of two-fluid hydrodynamics. At low but non zero temperatures ($T\ll\Delta$, $T\ll T_c$), the
root $u_{\mathrm{s1}}$ acquires an imaginary part exponentially small in temperature,
$\textrm{Im}\, u_{s,0}(T)\propto e^{-\Delta'/T}$, with an activation energy $\Delta'$
strictly larger than $\Delta$ \cite{Orbach1981}. This is because the fermionic quasiparticles of energy
$\Delta$ have zero group velocity, and thus cannot contribute to the damping.
Our results for this imaginary part are in agreement with Refs. \cite{Orbach1981,Kurkjian2017-2}
and with Landau roton-phonon theory \cite{Kurkjian2017-3}.
The collective mode also acquires a velocity shift $\delta c_{s,0}(T)=\textrm{Re}\, u_{s,0}(T)-c_{s,0}(0)$.
In the weak-coupling BCS limit, we agree with Kulik \textit{et al.} \cite{Orbach1981}
who predicted an exponentially small increase of the velocity:
\be
\delta c_{s,0}(T)\stackrel[T\to0]{1/k_{F}a\to-\infty}{=}\frac{2v_{\rm F}}{5\sqrt{3}}\sqrt{\frac{2\pi T}{\Delta}}e^{-\Delta/T}.
\ee
As shown in Fig. \ref{fig:svelsLT}, we find that after this exponential increase,
the velocity passes through a shallow maximum and then decreases.
This behavior is reminiscent of what Ref.~\cite{Escobedo} obtained with a low-energy effective theory.
On the contrary, in the BEC regime, we find that the velocity shift is always negative.

\section{Behavior near the critical temperature}
\label{sec:Tc}
In contrast with the low-temperature regime, the behavior of the collectives branches
near $T_c$ remains a controversial problem. The available predictions
neatly contradict themselves: Popov and Andrianov \cite{Popov1976} find the pure imaginary
dispersion relation
\be
\omega_q=-i\frac{7\zeta(3)v_{\rm F}q}{6\pi^3T_c} \bb{\sqrt{4\Delta^2+v_{\rm F}^2q^2}+2\Delta}, \label{Popov}
\ee
which indicates that $u_s(T)$ has a critical exponent of $1/2$, that is,
$u_s(T)\underset{(T_c-T)\to0}{\sim}  a\Delta \sim a'(T_c-T)^{1/2}$ with $a,a'\in i\mathbb{R}$.
In contradiction with this result, Ohashi and Takada \cite{Takada1997}
predict a real speed of sound with a critical exponent of $1/6$, that is,
$u_s(T)\underset{(T_c-T)\to0}{\sim} b(T_c-T)^{1/6}$, $b\in\mathbb{R}$.
These two studies are limited to the weak coupling regime $1/k_{\rm F}a\to-\infty$.
More recent studies dealing with the strong coupling regime \cite{Kosztin} confirmed the cancellation of the speed of sound
at $T_c$ (irrespectively of the interaction regime) but did not predict its critical exponent.
Using our dispersion equation \eqref{Wu}, we are in a good position to solve this controversy.

Using the mean-field equation of state (or the ``scaled GPF'' scheme described in Appendix
\ref{app:eos}, which preserves this limiting behavior), the limit $T\to T_c$ implies
\bea
\epsilon&\equiv&\frac{\Delta}{T} = O(\sqrt{T_c-T}), \\
\frac{\mu}{T} &=&\frac{\mu(T_c)}{T_c}  + O(T_c -T).
\eea
Neglecting terms of order $\epsilon^2$, we thus take the limit $\epsilon\to0$ for $\mu/T$ fixed to $m_c\equiv{\mu(T_c)}/{T_c}$.
Note that $m_c$ is related to $k_{\rm F}a$ by an equation of state at $T_c$, as explained in section \ref{sec:eos}.
This relation is of course different for, e.~g., the mean-field or scaled GPF equations of state.

\subsection{Regimes with ($\mu>0$)}

\begin{figure}[tbh]%
\includegraphics[width=3.5in]{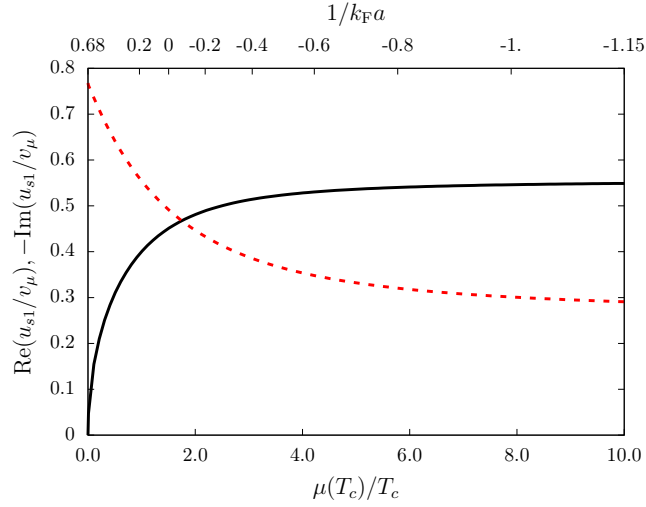}
\caption{(Color online) The first root $u_{s1}$ of the dispersion equation \eqref{dispTc} is plotted, in units of $v_\mu=\sqrt{2\mu/m}$, as a function of ${\mu(T_c)}/{T_c}$ (related by the equation of state to the interaction strength $1/k_{\rm F}a$, shown on the top $x$-axis)}%
\label{fig:u1}
\end{figure}

When $\mu(T_c)>0$ (that is for $1/k_{\rm F}a<0.68$ with the mean-field equation-of-state), the $m_{\sigma\sigma'}$ coefficients in the limit $\epsilon\to0$
become\footnote{The subleading terms  in $\tilde{m}_{++}$ and $\tilde{m}_{--}$ depend a priori on $\check{u}$ (see Appendix \ref{app:Tc}).
Since these terms matter only when $\check{u}^2F$ and $G$ are $O(\epsilon)$, that is when $\check{u}=O(\epsilon)$, we give
in (\ref{mppTc}--\ref{mpmTc}) only the value of these functions in $\check{u}=0$.}
\bea
\frac{\check{m}_{++}}{m_c}&=&\check{u}^2{F(\check{u})} +\epsilon f(m_c) +O(\epsilon^2) \label{mppTc} \\
\check{m}_{--} &=&\epsilon \bbcro{G(\check{u})+\epsilon g(m_c)}+O(\epsilon^3) \label{mmmTc} \\
\check{m}_{+-}&=&{\epsilon} h(m_c) +O(\epsilon^2) \label{mpmTc}
\eea
Since $\mu$ is the most convenient energy scale near $T_c$, we have redimensionalized the speed of sound, $\check{u}^2=mu^2/2\mu$, and the integrals in consequence, $m_{\sigma\sigma'}={\rho(\mu)\Delta}\check{m}_{\sigma\sigma'}/{2}$, where $\rho(\mu)=\sqrt{2m^3\mu}/\pi^2\hbar^3$ is the density of
states at energy $\mu$ (setting the volume of the gas equal to unity). We also introduced the functions
\bea
F(\check{u}) &=&-\frac{\pi}{8}\bb{\sqrt{1-\frac{1}{\check{u}^2}}+\check{u}\,\textrm{arccsc}(\check{u})} \label{F} \\
G(\check{u}) &=&\frac{\pi}{4}{\check{u}\,\textrm{arccsc}(\check{u})} \label{G}
\eea
and we recall that $\textrm{arccsc}(z)=-i\textrm{ln}\bb{\sqrt{1-\frac{1}{z^2}}+\frac{i}{z}}$. Functions $f$, $g$ and $h$ of $m_c$ are defined in Appendix \ref{app:Tc} where the
derivation is detailed. The dispersion equation \eqref{Wu} on $\check{u}$ then becomes
\be
\bbcro{{\check{u}^2}F(\check{u}) +\epsilon f(m_c) }\bbcro{G(\check{u})+\epsilon g(m_c)}-4{\check{u}^2}h^2(m_c)=0 \label{dispTc}
\ee
This equation should be solved in the lower-half complex plane after analytic continuation of the functions $F$ and $G$. With the analytic formulas Eqs.~(\ref{F},\ref{G}), this is simply done by the replacements $\sqrt{1-{1}/{\check{u}^2}}\to-\sqrt{1-{1}/{\check{u}^2}}$ and $\textrm{arccsc}(\check{u})\to\pi-\textrm{arccsc}(\check{u})$. Remarkably, we find that the analytically continued equation has in fact two solutions. The first one (shown in Fig.~\ref{fig:u1} as a function of $m_c$ or $1/k_{\rm F}a$) has a nonzero limit when $\epsilon\to0$; it is given by the transcendent equation
\be
F(\check{u}_{s1})G(\check{u}_{s1})=4h^2(m_c)
\ee
The second solution $u_{s2}$ behaves as $\epsilon\propto(T_c-T)^{1/2}$, which confirms the $1/2$ critical exponent predicted by Andrianov and Popov. Setting $\check{u}_{s2}=\epsilon\,\bar{u}_{s2}$, and simplifying Eq.~\eqref{dispTc} for $\epsilon\ll1$ (but $|\textrm{ln}\,\epsilon|=O(1)$) we obtain
\be
\bb{-i\frac{\pi\bar{u}_{s2}}{8} + { f(m_c)}}\bb{\vphantom{\frac{\check{u}^2}{2}} \frac{\pi\bar{u}_{s2}}{4}\bbcro{\pi+i\textrm{ln}\frac{i\bar{u}_{s2}\epsilon}{2}}+g(m_c)}-4{\bar{u}_{s2}^2}h^2(m_c)=0
\ee
Thus, $\bar{u}_{s2}$ still depends logarithmically on $\epsilon$. This dependence can in turn be expanded at temperatures extremely close to $T_c$, that is for $|\textrm{ln}\,\epsilon|\gg1$:
\be
\check{u}_{s2}=\epsilon \frac{8f(m_c)}{\pi} \bb{-i\bbcro{1+\frac{128h^2(m_c)}{\pi^2|\textrm{ln}\epsilon|}}+\frac{\gamma}{\textrm{ln}^2\epsilon}} +O\bb{\frac{\epsilon}{\textrm{ln}^3\epsilon}}
\ee


The first two terms of this expansion are pure imaginary numbers, while the term in $O(\epsilon/\textrm{ln}^2\epsilon)$ has a non-zero real part. The quality factor $\Re{u}_{s2}/2\Im{u}_{s2}$ thus vanishes near $T_c$ as $\gamma/2\textrm{ln}^{2}\epsilon$, where the coefficient $\gamma$ is:
\be
\gamma=-\frac{2^{12}h^2}{\pi^4} \bbcro{i\frac{\pi}{8}\bbaco{g+\frac{2^8}{\pi^2}h^2f-2if\bb{\pi+i\ln\frac{4f}{\pi}}}-64i\frac{h^2f}{\pi}}
\ee
with the short-hand notation $f=f(m_c)$, $g=g(m_c)$ and $h=h(m_c)$.

In the BCS limit ($m_c\to+\infty$ or $1/k_{\rm F}a\to -\infty$), one has $h=0$ in \eqref{dispTc}, such that the dispersion equation becomes $m_{++}m_{--}=0$, and $u_{s1}$ and $u_{s2}$ solve $m_{++}=0$ (while $m_{--}=0$ gives the pair-breaking Popov-Andrianov-``Higgs'' mode \cite{Popov1976,KurkjianPB}). Using the limiting value $f(+\infty)=7\zeta(3)/12\pi^2$, we get an expression of ${u}_{s2}$ that agrees with Andrianov-Popov, Eq.~\eqref{Popov}:
\be
\frac{u_{s2}}{v_{\rm F}}\stackrel[\epsilon\to0]{1/k_{\rm F}a\to -\infty}{=}-i\frac{14\zeta(3)}{3\pi^3}\frac{\Delta}{T_c},
\ee
Conversely, ${u}_{s1}$ has the finite nonzero limit
\be
\frac{u_{s1}}{v_{\rm F}}\stackrel[\epsilon\to0]{1/k_{\rm F}a\to -\infty}{\simeq}0.555 - 0.266 i
\ee

\begin{figure}[tbh]%
\centering
\includegraphics[width=3.5in]{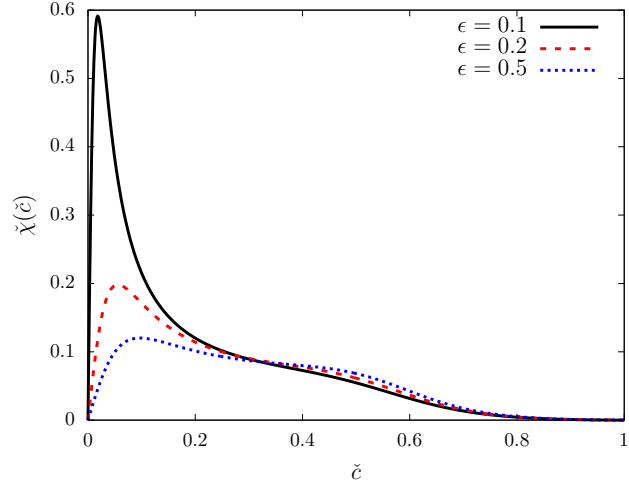}
\caption{(Color online) The phase-phase response function $\chi$ is plotted as a function of the reduced velocity $\check{c}=c\sqrt{m/2\mu}$ in
the far BCS regime, at $1/k_F a=-2$ and temperatures $T/T_c=0.999$ ($\epsilon=0.1$), $T/T_c=0.996$ ($\epsilon=0.2$),
and $T/T_c=0.97$ ($\epsilon=0.5$). Far from $T_c$, the two roots $\check{u}_{s1}$ and $\check{u}_{s2}$
of the dispersion equation have comparable imaginary parts (and comparable residues),
which results in a response function with a double bump structure (blue curve).
As the temperature is reduced, $\check{u}_{s2}$, whose real and imaginary
part tend to $0$ like $\epsilon$, dominates, which results in the large resonance peak near $c=0$ (black curve).
The contribution of $\check{u}_{s1}$ still leads to a shoulder at larger $c$.
}%
\label{fig:repTc}
\end{figure}

The existence of two solutions to the speed-of-sound equation, and thus of two phononic branches, is surprising but it is not an artifact of our analytic continuation scheme. It is confirmed by looking at the response function $\chi(c)$, which is a physical observable. Expressions (\ref{mppTc}-\ref{mpmTc}) can be used to express the response function near $T_c$:
\be
\check{\chi}(\check{c})=\frac{1}{\pi m_c}\Im\frac{G(\check{c})+\epsilon g}{(\check{c}^2F(\check{c})+\epsilon f)(G(\check{c})+\epsilon g)-4\check{c}^2h^2} \label{chiTc}
\ee
where the redimensionalization is $\check{\chi}={\chi}\times\bbcro{\rho(\mu)\Delta/2}$.
In Fig.~\ref{fig:repTc}, we show this response function in the far BCS regime $1/k_{\rm F }a=-2$ (corresponding to $\mu(T_c)/T_c\simeq37.73$). The second root, whose quality factor diverges when $T\to T_c$, translates into a sharp resonance peak whose center tends to $c=0$ and whose width vanishes at $T_c$. The first root, which conversely has a finite quality factor, does not lead to the appearance of a second peak at temperatures close to $T_c$ (we shall see that it does at lower temperatures); it is nevertheless observable in the form of a broad upper shoulder that extends to higher $c$.

\subsection{BEC regime ($\mu<0$)}

In the BEC regime ($\mu(T_c)<0$), we obtain the following expansions of the $m_{\sigma\sigma'}$ integrals:
\bea
\check{m}_{++}(\check{u})&=&\epsilon\bbcro{\alpha_1(m_c)+\check{u}^2 \alpha_2(m_c)} +O(\epsilon^3) \label{mppTcBEC} \\
\check{m}_{--}(\check{u})&=&\epsilon^2\bbcro{\beta(m_c)+ \check{u}B(\check{u},m_c)} +O(\epsilon^3) \label{mmmTcBEC} \\
\check{m}_{+-}(\check{u})&=&\epsilon\gamma(m_c) +\epsilon^3C(\check{u},m_c)+O(\epsilon^4) \label{mpmTcBEC}
\eea
where the
redimensionalization
is the same as in the BCS regime with $\mu$ replaced by $|\mu|$. The functions $\alpha_1,\alpha_2,\beta$ and $\gamma$ of $m_c$ are defined in Appendix \ref{app:Tc}, and the function $B$ is given by an integral
\be
B(\check{u},m_c)=\int_{|m_c|}^\infinity de \frac{ \textrm{tanh}'\bb{ e/2}}{4e^2} \textrm{arctanh}\bbcro{\frac{\sqrt{e/|m_c|-1}}{\check{u}}} \label{B}
\ee
We introduce the function $C(\check{u},m_c)$ for the sake of completeness,
but it is not needed to derive the speed of sound to leading order.
The dispersion equation \eqref{Wu} in the BEC regime near the transition temperature becomes
\be
\epsilon^2\bbcro{\alpha_1(m_c)+\check{u}^2 \alpha_2(m_c) }\bbcro{\beta(m_c)+ \check{u}B(\check{u},m_c)}-4 m_c \bbcro{\gamma(m_c)+\epsilon^2 C(\check{u},m_c)}^2\check{u}^2+O(\epsilon^3)=0 \label{dispTcBEC}
\ee
The analytic continuation of this  equation is only slightly more difficult than in the BCS regime; replacing in Eq.~\eqref{B} $\textrm{arctanh}(z)$ by $i\pi+\textrm{arctanh}(z)$ for $\Re z>1$, we obtain the analytic continuation $B_\downarrow$ of $B$:
\be
B_\downarrow(\check{u},m_c)=
\begin{cases}
B(\check{u},m_c) \quad\textrm{if}\quad\Im\,z>0 \\
B(\check{u},m_c) -i\pi\int_{(\Re\,\check{u}+1)^2|m_c|}^{+\infty}\frac{ \textrm{tanh}'\bb{ e/2}}{4e^2} de \quad\textrm{if}\quad\Im z<0 \\
\end{cases}
\ee
Note that $B_\downarrow(0,m_c)=-i\pi\int_{|m_c|}^{+\infty}\frac{ \textrm{tanh}'\bb{ e/2}}{8e^2} de$ is a pure imaginary number.
The analytically continued equation \eqref{dispTcBEC} admits a
single complex root $ \check{u}_{s,B}$
which tends to $0$ when $\epsilon\to0$. Up to order $\epsilon^2$, we can then neglect
the terms controlled by $\alpha_2$ and $C$ in \eqref{dispTcBEC}, to obtain:
\bea
\Re \check{u}_{s,B} &=& \epsilon\sqrt{\frac{\alpha_1\beta}{4|m_c|\gamma^2}}+O(\epsilon^3) \\
\Im \check{u}_{s,B} &=& \epsilon^2\frac{\alpha_1 B_\downarrow(0,m_c)}{8|m_c|\gamma^2} +O(\epsilon^3)
\eea
Contrarily to the BCS regime, there is here no remaining logarithmic dependence of $\check{u}_{s,B}/\epsilon$. Moreover,
the quality factor $\Re \check{u}_{s,B}/2\Im \check{u}_{s,B}$, instead of being logarithmically cancelled, now diverges like $1/\epsilon$.

Finally, in the BEC limit ($m_c\to-\infty$ or $1/k_{\rm F}a\to+\infty$), we use the equivalents
$
\alpha_1  \sim  |m_c|\beta\sim-\gamma/2  \underset{|m_c|\to+\infty}{\sim}  {\pi}/{16|m_c|} $ and
 $B_\downarrow(0,m_c)\underset{|m_c|\to+\infty}{\sim}  
-{i\pi e^{-|m_c|} }/{2|m_c|^2}
$
to obtain
\bea
\Re \check{u}_{s,B} &\stackrel[|m_c|\to+\infty]{\epsilon\to0}{\sim}& \frac{\epsilon}{4|m_c|} \label{usB}\\
\Im \check{u}_{s,B} &\stackrel[|m_c|\to+\infty]{\epsilon\to0}{\sim}& - \frac{i\epsilon^2}{4|m_c|^2}e^{-|m_c|} \label{kappasB}
\eea
The quality factor of the branch thus diverges exponentially with $|m_c|$ in the BEC limit. The damping of the collective modes
by the unpaired fermions becomes less efficient when the pairs form a weakly interacting condensate of dimers.
Note that our results may be
less meaningful in the BEC limit where one expects purely bosonic effects not
described by GPF, such as phonon-phonon couplings, to play a major role. It is
known for example that important corrections to $T_{c}$ arise when taking into
account the condensate depletion due to the bosonic branch \cite{deMelo1993}.

\section{Numerical results at intermediate temperatures}
\label{sec:analcont}

\subsection{Numerical method for the analytic continuation}

When the temperature is neither close to $0$ nor to $T_c$,
it is impossible to express the dispersion equation with simple analytic
formulas such as \eqref{dispTc} or \eqref{dispTcBEC}, and thus to perform the analytic
continuation based on the analytic properties of elementary functions. We thus
develop a numerical method based on the procedure of Nozi\`eres \cite{Nozieres,KurkjianPB},
which is able to perform the analytic continuation
directly from the integral expression Eqs.~(\ref{mpp}--\ref{mpm}).

\paragraph{Spectral functions}
Quite generally, we consider a function $F$ of the complex variable
$ u$ having a branch cut at the real axis for $u=c\in
\mathbb{R}$, and introduce the associated
spectral function,%
\begin{equation}
\rho_{F}\left(  c\right)  =-\lim_{\delta\rightarrow0}\frac{F\left(
c+i\delta\right)  -F\left(  c-i\delta\right)  }{2\pi i}. \label{rho}%
\end{equation}
The spectral function $\rho_{F}\left(  c\right)  $ is in general analytic on
the real axis except at most on a finite number of points. It can thus be
analytically continued from any chosen interval between these points to the
lower complex half-plane. The analytic continuation $F^{\left(  I\right)
}\left(  u\right)  $ of $F\left(  u\right)  $ from upper to lower complex
half-plane and through the interval $I\subset \mathbb{R}$ where $\rho_{F}$ is analytic
then reads:%
\begin{equation}
F^{\left(  I\right)  }\left(  u\right)  =\left\{
\begin{array}
[c]{cc}%
F\left(  u\right)  , & \operatorname{Im}u>0,\\
F\left(  u\right)  -2\pi i\rho_{F}^{(I)}\left(  u\right)  , &
\operatorname{Im}u<0,
\end{array}
\right.  \label{ancont}%
\end{equation}
where $u\mapsto\rho_{F}^{(I)}\left(  u\right)  $ is the analytic continuation
of $\rho_{F}\left(  c\right)  $ from the interval $I$ to the lower complex half-plane.

To perform the analytic continuation of functions $m_{\sigma,\sigma'}$, we
compute their spectral functions and study their singularities on the real axis.
After the angular integration over $\theta$ in Eqs.~(\ref{mpp}--\ref{mpm}), we get:
\begin{align}
m_{++}(u)  & =\int_0^\infty \frac{\Delta^{2}k^{2}%
dk}{2\pi^2}~\left[  \frac{mv_k^{2}}{12E_{k}^{2}}\left(  \frac{X}{E_{k}%
}-X^{\prime}\right)  -\frac{mu^{2}}{4E_{k}^{2}}\left(  \frac{X}{E_{k}}%
+\frac{\Delta^{2}}{\xi_{k}^{2}}X^{\prime}\right)  \right.  \nonumber\\
& \left.  +\frac{\Delta^{2}X^{\prime}mu^{3}}{8\xi_{k}^{3}E_{k}v_k}\bbaco{ \ln\left(  uE_{k}+\xi_{k}v_k\right)-\ln\left(
uE_{k}-\xi_{k}v_k\right)}  %
\right]  ,\label{mpp2}\\
m_{--}(u)  & =\int_{0}^{\infty}\frac{\Delta^{3}k^{2}d{k}}{2\pi^{2}}\left[
\frac{1}{2E_{k}^{2}}\left(  \frac{X}{E_{k}}-X^{\prime}\right)  +\frac
{{u}X^{\prime}}{4{E}_{k}{\xi}_{k}v_{k}}\left\{  \ln(E_{k}{u}+v_{k}{\xi}%
_{k})-\ln(E_{k}{u}-v_{k}{\xi}_{k})\right\}  \right]  ,\label{mmm2}\\
m_{+-}(u)  & =\int_{0}^{\infty}\frac{\Delta^{2}k^{2}d{k}}{2\pi^{2}}\left[
-\frac{X{\xi}_{k}}{4{E}_{k}^{3}}-\frac{X^{\prime}}{4{E}_{k}^{2}{\xi}_{k}%
}+\frac{{u}\Delta^{2}X^{\prime}}{8{E}_{k}{\xi}_{k}^{2}v_{k}}\left\{  \ln
(E_{k}{u}+v_{k}{\xi}_{k})-\ln(E_{k}{u}-v_{k}{\xi}_{k})\right\}  \right]
,\label{mpm2}%
\end{align}
with the short-hand notations $X=X(E_k)$ and $X'=X'(E_k)$. In these expressions,
the only contribution to the spectral functions comes from the logarithms
that have a discontinuity $\ln(x+i0^+)-\ln(x-i0^+)=2i\pi$ for $\Re x<0$.
We then obtain generically
\begin{equation}
\rho_{\sigma\sigma'}(c)= {c}^p\left[  \int_{I_{-}(c)}dkf_{\sigma\sigma'}(k)-\int_{I_{+}(c)}%
dkf_{\sigma\sigma'}(k)\right]  \label{Fc}%
\end{equation}
where $p=1$ for $\rho_{--}$ and $\rho_{+-}$ and $p=3$ for $\rho_{++}$.
Note that the integrands $f_{\sigma\sigma'}$ (whose exact expressions follow immediately from
Eqs.~(\ref{mpp2}--\ref{mpm2})) are independent of $c$, such that the only dependence on $c$ (besides the trivial prefactor)
is through the integration intervals $I_{\pm}(c)$. The idea of our numerical method is to compute analytically
the boundaries of those intervals, which we then analytically continue to the complex plane, yielding the continuations
of the spectral functions $\rho_{\sigma\sigma'}(u)$, $u\in\mathbb{C}$.

\begin{figure}[tbh]%
\centering
\includegraphics[
width=3.5in
]%
{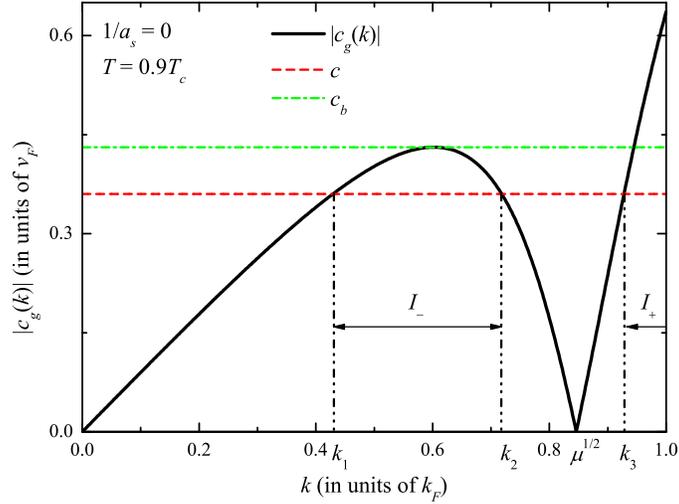}%
\caption{(Color online) \textit{Solid curve}: the absolute value of the group
velocity for the Bogoliubov excitations at $1/k_F a=0$ and $T=0.9T_{c}$.
\textit{Dot-dashed line}: the boundary velocity
$c_{b}$ at the same parameters. \textit{Dashed line:} an example of
a value of $c$ at which the spectral function is computed. When $c<c_{b}$, the integration interval in
the spectral function is made of two disconnected intervals whereas for
$c>c_{b}$, it is made of a single interval.}%
\label{fig:VG}%
\end{figure}

\paragraph{Resonance intervals}
The intervals $I_{\pm}(c)$ are defined as the set of wave numbers $k$ where the argument
of the logarithms in Eq.~(\ref{mpp2}--\ref{mpm2}) has a negative real part, which leads to the condition
\begin{equation}
c<\pm c_{g}\left(  k\right)  , \label{crit}%
\end{equation}
Here, $c_{g}\left(  k\right)  =\frac{\partial E_{{k}}}{\partial
k}=k\xi_{{k}}/mE_{{k}}$ is the group velocity of the BCS
fermionic excitations. This velocity is positive for $k>\sqrt{2m\mu}$ and
negative for $0<k<\sqrt{2m\mu}$; it is represented in absolute value in Fig.~\ref{fig:VG}. In Refs.~\cite{Kurkjian2017-2,Zhang} condition \eqref{crit}
was derived as the low-$q$ version of the resonance condition $\omega
_{\mathbf{q}}=E_{\mathbf{k}+\mathbf{q}}-E_{\mathbf{k}}$ after angular
integration. In Ref. \cite{Kurkjian2017-2} it has been further interpreted as
a Landau criterion considering an unpaired fermion as an impurity moving in the superfluid.%

Since $c_{g}(k)\rightarrow\infty$ when $k\rightarrow\infty$, the inequality
$c<c_{g}\left(  k\right)  $  {is} always fulfilled for large
enough $k>\sqrt{2m\mu}$. The interval $I_{+}(c)$ is then of the form
$[k_{3}(c),+\infty\lbrack$. As visible in Fig.~\ref{fig:VG}, the inequality
$c<-c_{g}\left(  k\right)  $ can be also fulfilled at lower $k$ ($0<k<\sqrt
{2m\mu}$) provided that $c$ is small enough, that is smaller than the boundary
velocity,%
\begin{equation}
c_{b}=\sqrt\frac{{2\mu+3\Delta\bb{\frac{1}{s^{1/3}}-s^{1/3}}}}{{m}},\quad s\equiv\left(  \mu/\Delta+\sqrt{\mu^{2}/\Delta
^{2}+1}\right)  , \label{v0}%
\end{equation}
which is the absolute value of the minimum of the group velocity, $c_{b}%
=|\min_{k}\left[  c_{g}\left(  k\right)  \right]  |$ (in other words, the
largest slope of the BCS branch $k\mapsto E_{k}$ in its decreasing part). The
boundary sound velocity $c_{b}$ decreases when moving from the BCS to the BEC
regime and vanishes when $\mu=0$, that is when the
decreasing part of the BCS branch disappears. At a fixed scattering length,
$c_{b}$ rises with increasing temperature because the chemical potential
$\mu(T)$ rises. When the
condition $c<c_{b}$ is fulfilled, the interval $I_{-}(c)$ exists and is of the
form $[k_{1}(c),k_{2}(c)]$. Since $c_{g}(\sqrt{2m\mu})=0$, when the two momentum
ranges exist they are disjoint ($k_{3}(c)>k_{2}(c)$). The boundary functions
$k_{j}\left(  c\right)  $, $j=1,2,3$, when they exist, are the real positive roots of
the polynomial equation,%
\begin{equation}
\xi_{{k}}^{3}+\left(  \mu-\frac{mc^{2}}{2}\right)  \xi_{{k}}%
^{2}-\frac{mc^{2}}{2}\Delta^{2}=0, \label{polynome}
\end{equation}
with $\xi_{{k}}=k^{2}/2m-\mu$.

When $c\rightarrow c_{b}$ from below, the integral over $I_{-}$ in \eqref{Fc}
tends to $0$, but its derivative can remain finite, which results in an
angular point of the spectral function $\rho_{\sigma\sigma'}$ in $c=c_{b}$. Physically,
this angular point corresponds to the opening or closing of a decay channel in
the decreasing part of the BCS branch, at $k<\sqrt{2m\mu}$. This angular point
will become a branch point in the analytic continuation.

\paragraph{Choices for the analytic continuation}

\begin{figure}[tbh]%
\centering
\includegraphics[
width=3in
]{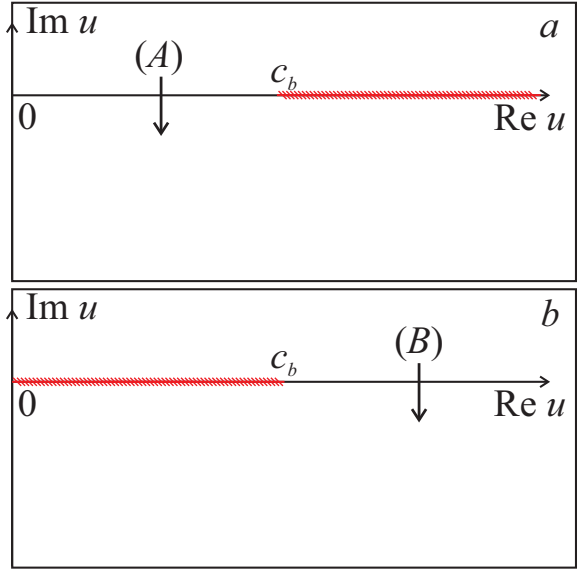}%
\caption{(Color online) Scheme of possible analytic continuations of
coefficients $m_{\sigma\sigma'}$ to the lower half-plane $\operatorname{Im}u<0$:
through \textquotedblleft window\textquotedblright\ $A$ (panel $a$),
\textquotedblleft window\textquotedblright\ $B$ (panel $b$). The branch cut
of the analytically continued functions are shown by striped lines. The
different analytic continuations have slightly different roots of the
dispersion equation $W_\downarrow\left(  u\right)  =0$.}%
\label{fig:branch}%
\end{figure}

The spectral functions $\rho_{\sigma\sigma'}\left(  c\right)  $ are analytic {separately} in
the interval {$A=[0,c_{b}[$} and {$B=]c_{b},+\infty\lbrack$}. Therefore there
are two possible ways to continue them to $\operatorname{Im}(u)<0$:
\begin{align}
\rho_{\sigma\sigma'}^{(A)}(u)  &  =u^p\bbcro{\int_{k_{3}(u)}^{+\infty}dkf_{\sigma\sigma'}(k,u)-\int_{k_{1}%
(u)}^{k_{2}(u)}dkf_{\sigma\sigma'}(k,u)},\label{rhoA}\\
\rho_{\sigma\sigma'}^{(B)}(u)  &  =u^p\int_{k_{3}(u)}^{+\infty}dkf_{\sigma\sigma'}(k,u), \label{rhoB}%
\end{align}
where $k_{j}\left(  u\right)  $ for $j=1,2,3$, are the analytic continuations
of the real solutions of \eqref{polynome}. Numerically, these continuations
are obtained by an adiabatic follow-up of the roots of \eqref{polynome} in the complex plane.
Note that $k_1$ and $k_2$ can be continued
to the entire half-plane with $\Im u<0$ even though they are real only in the interval $[0,c_b]$
of the real axis.
Choices (\ref{rhoA}) and (\ref{rhoB}) for the analytic
continuation of the spectral functions translate into two possible analytic
continuations $m_{\sigma\sigma'}^{\left(  A\right)  }\left(  u\right)  $ and $m_{\sigma\sigma'}^{\left(
B\right)  }\left(  u\right)  $. As shown in Fig.~\ref{fig:branch},
when the analytic continuation is performed through the \textquotedblleft
window\textquotedblright\ $A$, a branch cut remains on interval $B$, and
\textit{vice-versa}.

\subsection{Results and discussion \label{sec:results}}

Using our ``complex boundaries'' numerical method to perform the analytic continuation,
we study the solutions of the dispersion equation in the whole range $[0,T_c]$.
The existence of two roots near $T_c$ is confirmed by our numerical study,
a finding that does not depend on the choice of \textquotedblleft window\textquotedblright\ $A$ or $B$
for the analytic continuation.
In order to make the results quantitatively relevant for comparison with experiments,
the sound velocity and damping are calculated here using
the ``scaled GPF'' equation of state
described in Appendix \ref{app:eos}.

\begin{figure}[tbh]%
\centering
\includegraphics[
width=3.5in
]%
{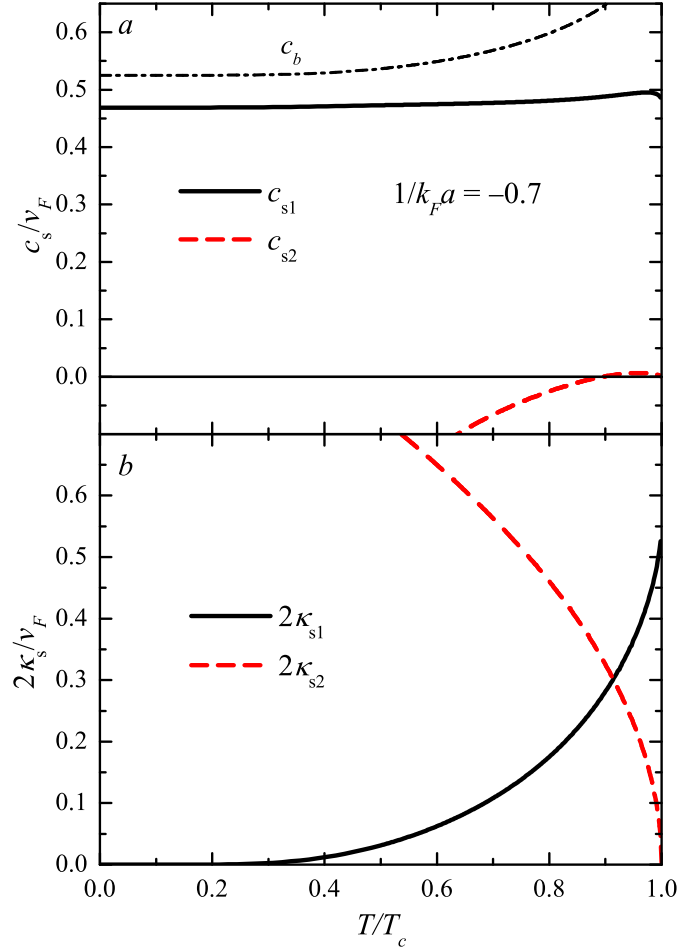}%
\caption{(Color online) The real (\textit{a}) and
imaginary parts $(b)$ of the two sound velocities
$u_{\mathrm{s,1}}$ (black solid curves) and $u_{\mathrm{s,2}}$ (red dashed curves)
in the BCS regime at $1/k_F a=-0.7$
(corresponding with the GPF equation of state to $  \mu(T_c)/T_c\simeq 6.06$
and $\left.  \mu/\Delta\right\vert _{T=0}\simeq 3.34$) as functions of $T/T_c$.
The dashed-dotted curve shows the boundary velocity $c_b$ (Eq.~\eqref{v0}) between
sector $A$ (below $c_b$) and $B$ (above $c_b$) of the real axis.}%
\label{fig:svBCS}%
\end{figure}

\begin{figure}[tbh]%
\centering
\includegraphics[
width=3.5in
]%
{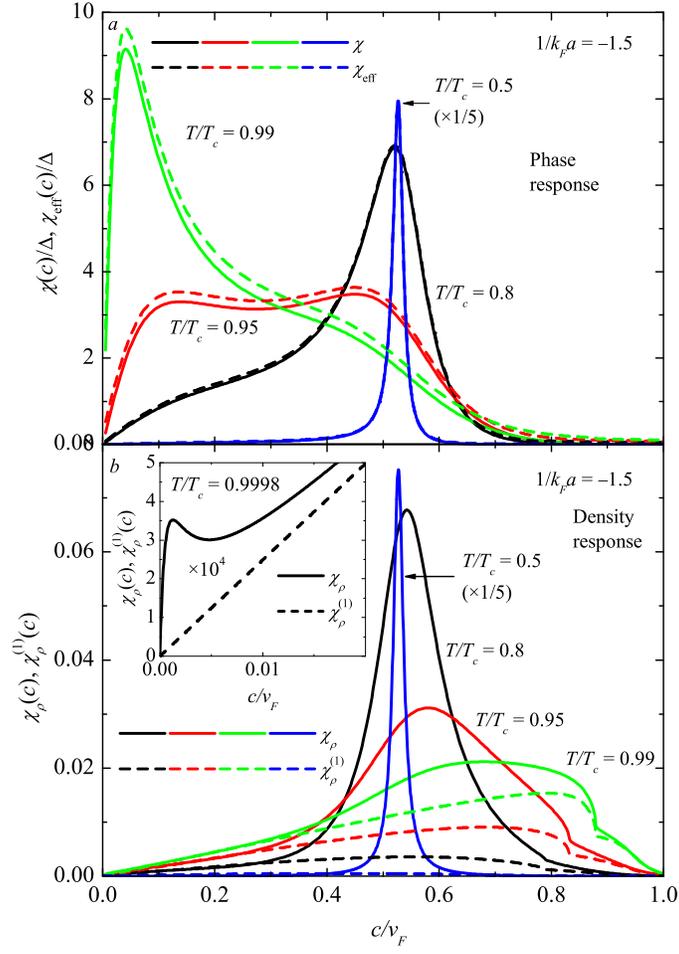}%
\caption{(Color online) Long-wavelength phase-phase response function
$\chi\left(  c\right)  $ (solid lines) and its two-pole analytic approximation
$\chi_{\mathrm{eff}}\left(  c\right)  $ (dashed lines) in the BCS regime ($1/k_Fa=-1.5$,
corresponding with the scaled GPF equation of state to $\left.  \mu/\Delta\right\vert _{T=0}\simeq 12.1$)
for $T=0.5T_{c}$ (blue lines), where they show a single quasi-Lorentzian peak,
$T=0.8T_{c}$ (black lines), where the peak is displaced and skewed by the increasingly
contributing second root, $T=0.95T_{c}$ where two resonances are visible (red lines)
and $T=0.99T_{c}$ (green lines) where the low-velocity resonance found in Sec.~\ref{sec:Tc} dominates.
(\emph{b}) Long-wavelength density-density response function $\chi_{\rho}\left(
c\right)  $ (solid curves) and the contribution of the pure density response
$\chi_{\rho}^{\left(  1\right)  }\left(  c\right)  $ (dashed curves) for the
same parameters of state as in the panel (\emph{a}).
\textit{Inset}: Low-velocity part of the long-wavelength density-density response function
in the close vicinity of the transition temperature, $T=0.9998T_{c}$.}
\label{fig:rep}%
\end{figure}

\paragraph{BCS and around unitarity regimes}
In the deep BCS regime, as shown in Fig.~\ref{fig:svBCS}
the speed of Anderson-Bogoliubov first sound $c_{s,0}$ found at zero temperature
evolves to the first root $u_{s,1}$ which we found near $T_c$.
Both its real and imaginary part $c_{s,1}$ and $\kappa_{s,1}$ are monotonically
increasing functions of temperature. The second solution $u_{s,2}$ appears only
above a threshold temperature\footnote{We define the threshold temperature $T_{th}$
as the temperature at which $\Re u_{s,2}$ reaches $0$. Below this temperature, the solution
$u_{s,2}$ still exists formally in the region of the complex plane with $\Re u_{s,2}<0$
(which nothing forbids our analytic continuation from accessing) but it has
little relevance for the response function.} $T_{th}$, which
tends to $T_c$ in the BCS limit. Its real part $c_{s,2}$ is zero at $T_{th}$ and at $T_{c}$ while its
imaginary part $\kappa_{s,2}$ monotonically decreases with temperature. There is thus a regime
in the range $[T_{th},T_{c}]$ where the two solutions are both well separated in frequency and
comparable in damping. As visible in Fig.~\ref{fig:rep},
the response function $\chi$ exhibits in this regime two distinguishable maxima
(not just a peak with a shoulder as in Fig.~\ref{fig:repTc})
corresponding to the two roots of the analytic continuation.
This unexpected finding is one of our key results; it
validates the existence
of two speeds of sound and thus of two collective modes in the GPF theory.

\begin{figure}[tbh]%
\centering
\includegraphics[
height=3.5in
]%
{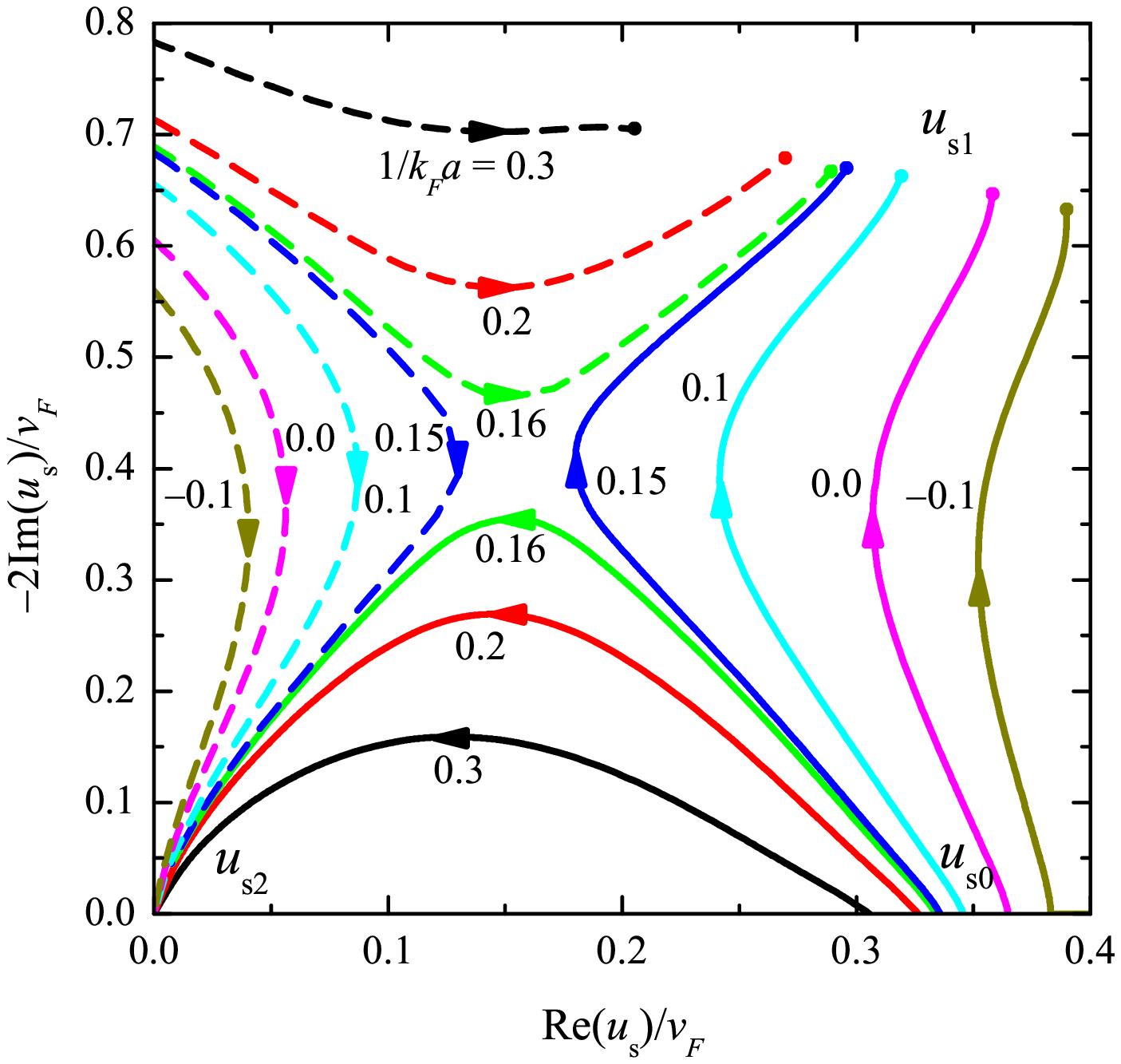}%
\caption{(Color online) The flow of the two roots of the dispersion equation \eqref{Wu} as a
function of temperature, for different values of the inverse scattering
length. The full curves show how the zero temperature root $c_{\mathrm{s}0}$ (lower right corner)
evolves to either $u_{\mathrm{s}1}$ (upper right corner) or $u_{\mathrm{s}2}$ (lower left corner)
as the temperature is increased up to $T_c$.
The dashed curves show the temperature evolution of the other root,
that appears in the upper left corner at the threshold temperature $T_{th}$.
The labels near each
curve show the corresponding value of $1/k_Fa$ for this curve. The curves are obtained
from the analytic continuation through window $A$, but the picture is qualitatively the same
using window $B$.}%
\label{fig:Paths}%
\end{figure}

At $1/k_F a = 1/k_F a_ {\mathrm{cross}}\simeq0.155$
(corresponding with the GPF equation of state to
$\mu(T_c)/T_c\simeq 1.376$, hence still in the non-BEC regime of Sec.~\ref{sec:Tc}),
an exact crossing of the two roots occurs at a given temperature:
$u_{\mathrm{s1}}(T_{\mathrm{cross}})=u_{\mathrm{s2}}(T_{\mathrm{cross}})$.
Then, for $1/k_Fa>1/k_Fa_{\mathrm{cross}}$, the situation changes: the zero
temperature solution $c_{s,0}$ evolves to $u_{s,2}$, while $u_{s,1}$ appears only
above the threshold temperature $T_{th}$. As illustrated in Fig.~\ref{fig:Paths},
this behavior is reminiscent
of that of two repulsive particles in $2D$, with temperature playing the role of time.
 The repulsion ensures that the
trajectories never cross: if the $x$-coordinates (here $\Re u$) cross,
then the $y$-coordinates (here $\Im u$) anticross, and
\textit{vice-versa}. In this analogy, the particular case $a=a_{\mathrm{cross}
}$ corresponds to the infinite energy case where the two particles exactly meet.

\paragraph{BEC regime}

As in Sec.~\ref{sec:Tc}, we define the boundary of the BEC regime as the point
where the chemical potential passes the zero value. Since $\mu$ depends on temperature,
the condition $\mu(T)=0$ corresponds to different values of the interaction
strength for different temperatures.
In particular, with the GPF equation of state $\mu(T=0)=0$ results in $1/k_F a \approx 0.427$, and $\mu(T_c)=0$ in $1/k_F a \approx 0.448$.
The corresponding values of the inverse scattering length with the mean-field equation of state are, respectively,
$1/k_F a \approx 0.553$ and $1/k_F a \approx 0.679$.

\begin{figure}[tbh]%
\centering
\includegraphics[
width=3.5in
]%
{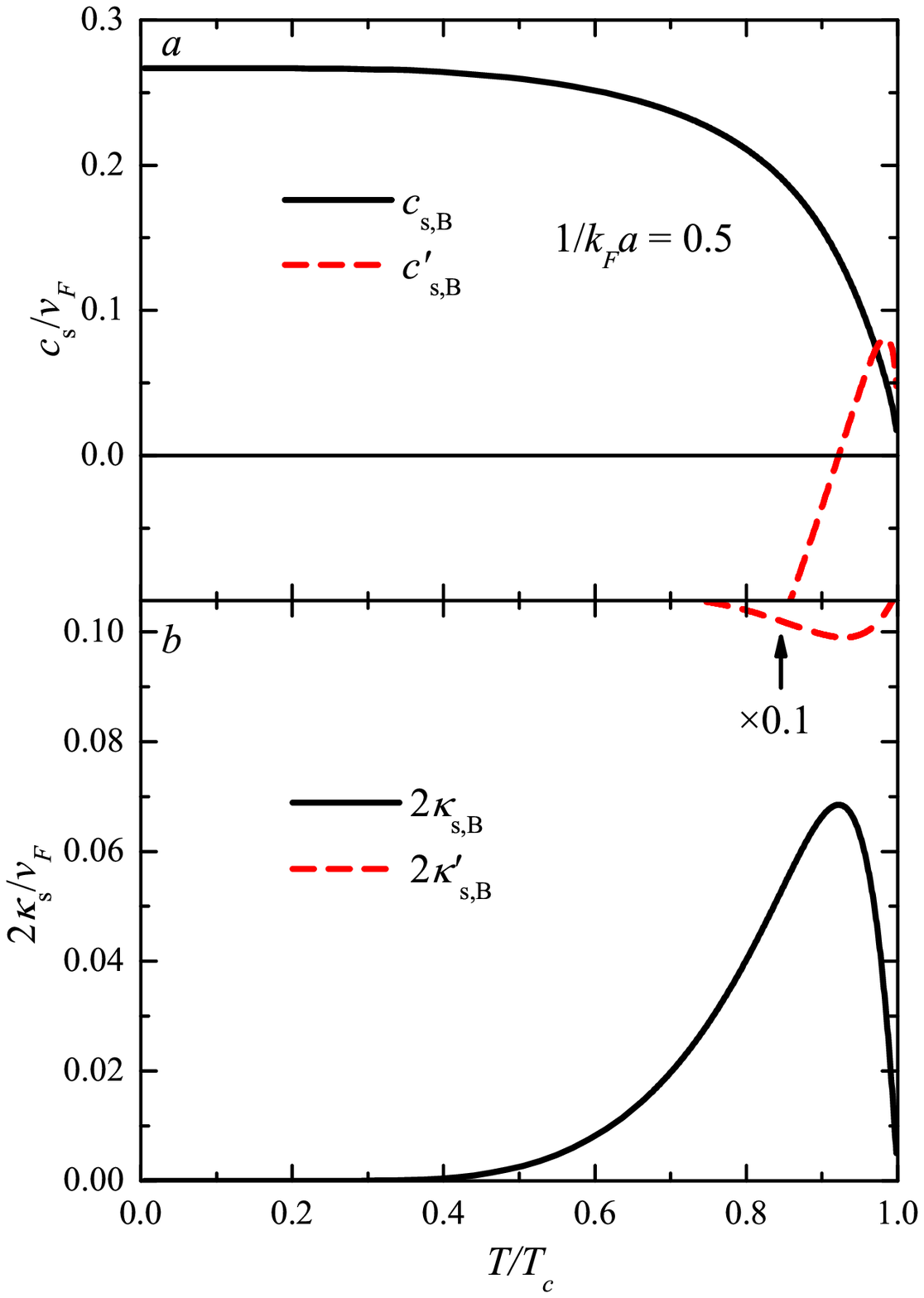}%
\caption{(Color online) The real (\textit{a}) and imaginary parts
(\textit{b}) of the BEC regime sound velocity (black solid line)
are shown at $1/k_Fa=0.5$
(corresponding with the GPF equation of state to $\mu(T_c)/T_c\simeq -0.293$
and $\left.  \mu/\Delta\right\vert _{T=0}\simeq -0.139$) as functions of $T/T_c$.
A second root (dashed red line) still exists in this regime
but it is highly damped and thus irrelevant for the response function.}%
\label{fig:svBEC}%
\end{figure}
%

In the BEC regime, represented in Fig.~\ref{fig:svBEC}, the $T=0$ solution $c_{s,0}$ always evolves to the solution
$u_{s,B}$ that we found near $T_c$. Its real part  $c_{s,B}$ decreases
monotonically with temperature, while its imaginary part  $\kappa_{\mathrm{s},B}$ vanishes at both  $0$ and $T_c$, and
goes through a maximum
in between. The height of this maximum
tends to zero in the BEC limit ($1/k_F a\to+\infty$), such that $\kappa_{\mathrm{s},B}(T)$
uniformly tends to zero in this limit. This is consistent with what we found
in the vicinity of $T_c$ (Eq.~\eqref{kappasB}) and indicates that the damping mechanism
we study (absorption-emission of collective excitations by fermionic quasiparticles)
becomes less relevant in the BEC limit where the condensed pairs weakly interact with the unpaired fermions.
As visible in Fig.~\ref{fig:svBEC}, a second solution still exists in the BEC regime, but it is
always largely damped such that it does not contribute to the response function,
which never displays the two-peak behavior we described in the BCS regime.

\paragraph{Visibility of the phase collective modes in the density response}

\begin{figure}[tbh]%
\centering
\includegraphics[
width=3.5in
]%
{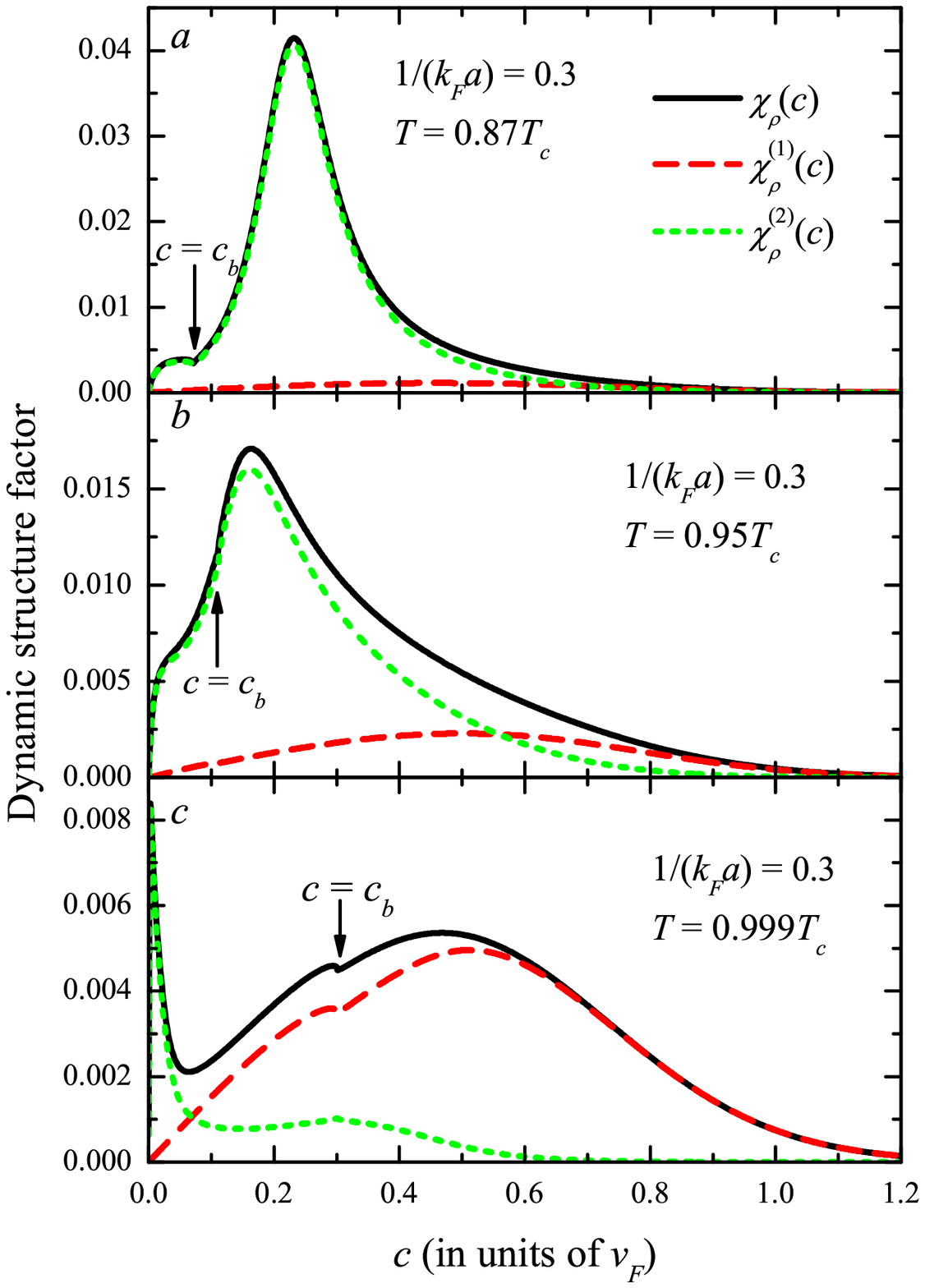}%
\caption{(Color online) Long-wavelength density response (Eq.~\eqref{lwlim}) function for $1/\left(
k_{F}a\right)  =0.3$ at (\emph{a}) $T=0.87T_{c}$, (\emph{b}) $T=0.95T_{c}$ and
at (\emph{c}) $T=0.999T_{c}$. Solid curves represent the total
density response. Dashed and dotted curves show, respectively, contributions
of pure density $  \chi_{\rho}^{\left(  1\right)  }  $ and order-parameter fluctuations
$  \chi_{\rho}^{\left(  2\right)  }  $ to
the total density response. The arrows indicate the boundary velocity $c_{b}$
determined by (\ref{v0}).}%
\label{fig:respdd}%
\end{figure}

The two phase collective mode we have found are also visible in the density-density response
function, as shown on Figs. \ref{fig:rep} (\emph{b}) and \ref{fig:respdd}. At low temperature
(blue curve in Fig. \ref{fig:rep} (\emph{b})),  the sole feature of $\chi_{\rho}\left(  c\right)$ (which is uniformly
dominated by the contribution $\chi_{\rho}^{(2)}$ of the pairing field)
is the Anderson-Bogoliubov resonance. When the temperature rises, the Anderson-Bogoliubov
resonance broadens and two other phenomena are visible: a broad incoherent peak due to the normal
component $\chi_{\rho}^{(1)}$ appears at velocities $c$ of order $v_{F}$.
This peak is not due to a collective mode (it does not have a Lorentzian shape)
but simply to the density response of a normal Fermi gas which becomes increasingly dominant near $T_c$.
At the same time, a resonance due to the $u_{s,2}$ pole of the
pairing-field propagator forms at low-velocities, and becomes increasingly sharp when $T\to T_c$.
The spectral weight of this resonance grows with increasing interaction strength (compare the inset of
Fig. \ref{fig:rep} (\emph{b}) in the BCS limit to Fig. \ref{fig:respdd} (c) at strong coupling).

Note that in the density response we can clearly see the boundary velocity $c_{b}$
discussed above, which is related to the opening of a decay channel in the
decreasing part of the BCS branch of excitations.%

\paragraph{Influence of the choice of the analytic continuation}

So far we have not discussed the physical consequences of
having two windows $A$ ($0\leqslant c\leqslant c_{b}$) and $B$ ($c\geqslant
c_{b}$) for the analytic continuation. For this, we go back to the physical
observable, which are the response functions $\chi$ and $\chi_\rho$.
Both of them have an angular point in $c_{b}$;
this is an observable feature, not an artifact of the approximation we have used, nor of the collisionless regime. In fact,
this angular point follows directly from energy
conservation, and is caused by the non-monotonic nature of the quasiparticle
spectrum which ensures that the low- and high-$k$ modes
are separated by a point of zero group-velocity:
the minimum of the BCS branch.
Thus, this angular point is in a sense a signature of the superfluid phase.
The physical meaning of the two windows $A$ and $B$ is then physically clear:
window $A$ is appropriate to reproduce the low-velocity ($c<c_{b}$) part of the response functions
and window $B$ for the high-velocity part ($c>c_{b}$).

When $c_b$ is far from the interesting features of the response function, that is from
the resonance peaks centered around $c_{s,1}$ and $c_{s,2}$, then only one restriction
of $\chi$, and thus only one analytic continuation, is worth studying.
This is the case for example in the BCS limit:
$c_b$ tends to $v_F$ which is well above both $c_{\mathrm{s},1}$ and $c_{\mathrm{s},2}$.
The \textquotedblleft window\textquotedblright\ $A$ (where the decay to
quasiparticle of wave number $k<\sqrt{2m\mu}$ is allowed) is then the
only choice. This
reflects the fact that the BCS branch has a large decreasing part  in this limit.
Similarly, in the BEC regime, one has $c_{b}=0$, so that only the \textquotedblleft window\textquotedblright\ $B$
is available for the analytic continuation.

\begin{figure}[tbh]%
\centering
\includegraphics[
width=3.5in
]%
{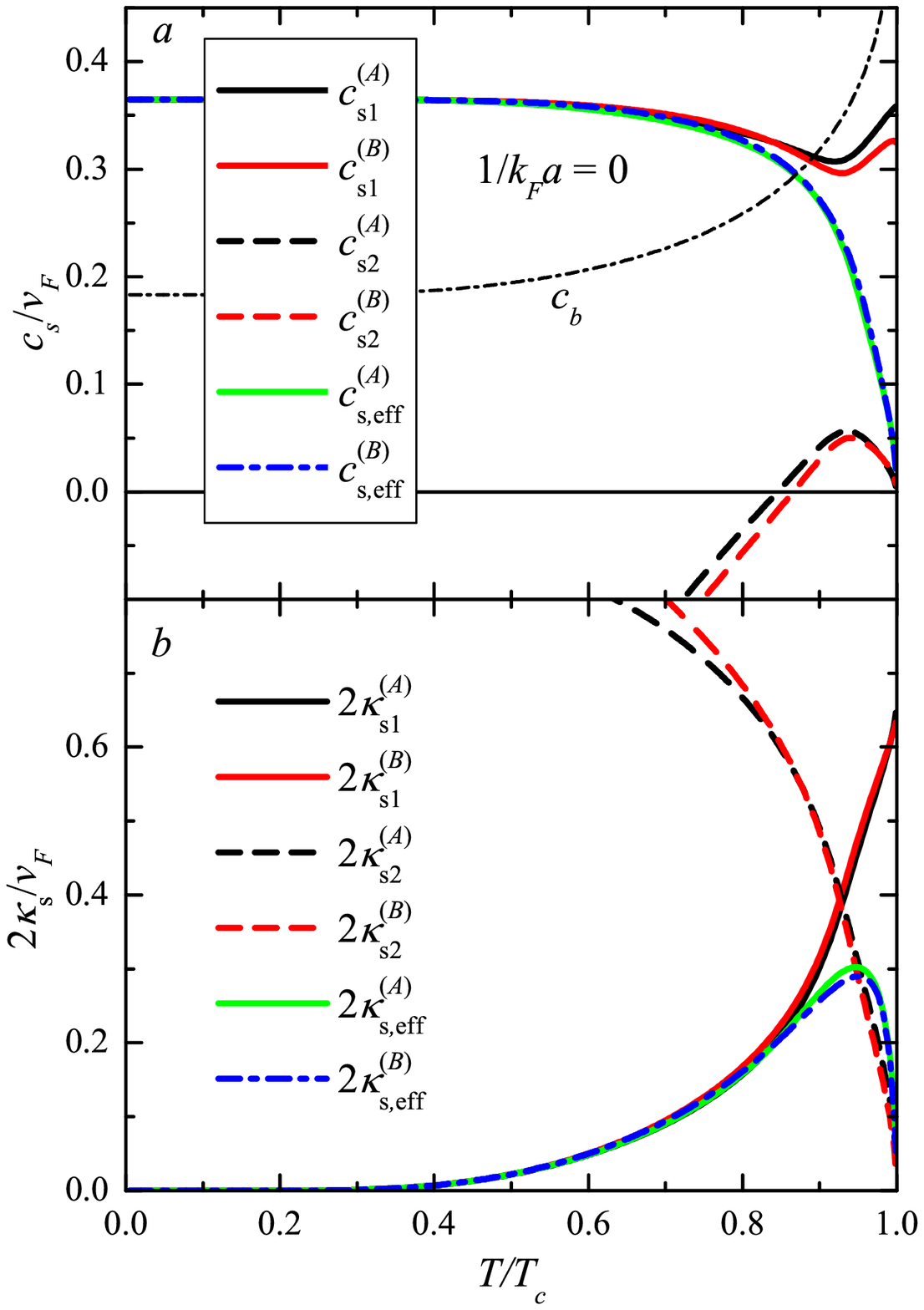}%
\caption{(Color online) The real (\textit{a}) and
imaginary parts $(b)$ of the two sound velocities
$u_{\mathrm{s,1}}$ (black and red solid lines) and $u_{\mathrm{s,2}}$ (black and red dashed lines)
are shown at unitarity $1/k_F a=0$ (corresponding to $  \mu(T_c)/T_c\simeq 1.50$
and $\left.  \mu/\Delta\right\vert _{T=0}\simeq 0.86$) as functions of $T/T_c$.
The dashed-dotted curve shows the boundary velocity $c_b$ (Eq.~\eqref{v0}) between
sectors $A$ and $B$ of the real axis. In this regime, the resonance gets close to $c_b$,
such that we should use window $A$ to describe the lower part
($0\leqslant c \leqslant  c_b$) of the resonance ($u_{\mathrm{s,1}}^{(A)}$ and $u_{\mathrm{s,2}}^{(A)}$ are shown in black),
and window $B$ for the upper part ($c_b \leqslant  c$, $u_{\mathrm{s,1}}^{(B)}$ and $u_{\mathrm{s,2}}^{(B)}$ are shown in red).
The green solid and blue dashed dotted lines show the effective sound velocities $u_{\mathrm{s,eff}}=c_{\mathrm{s,eff}}-i\kappa_{\mathrm{s,eff}}$ (defined in paragraph $d$ of Sec.~\ref{sec:results})
which characterize the position and width of the peak in the response function $\chi$, which is always unique at unitarity.}%
\label{fig:svUn}
\end{figure}

\begin{figure}[tbh]%
\centering
\includegraphics[
width=3.5in
]%
{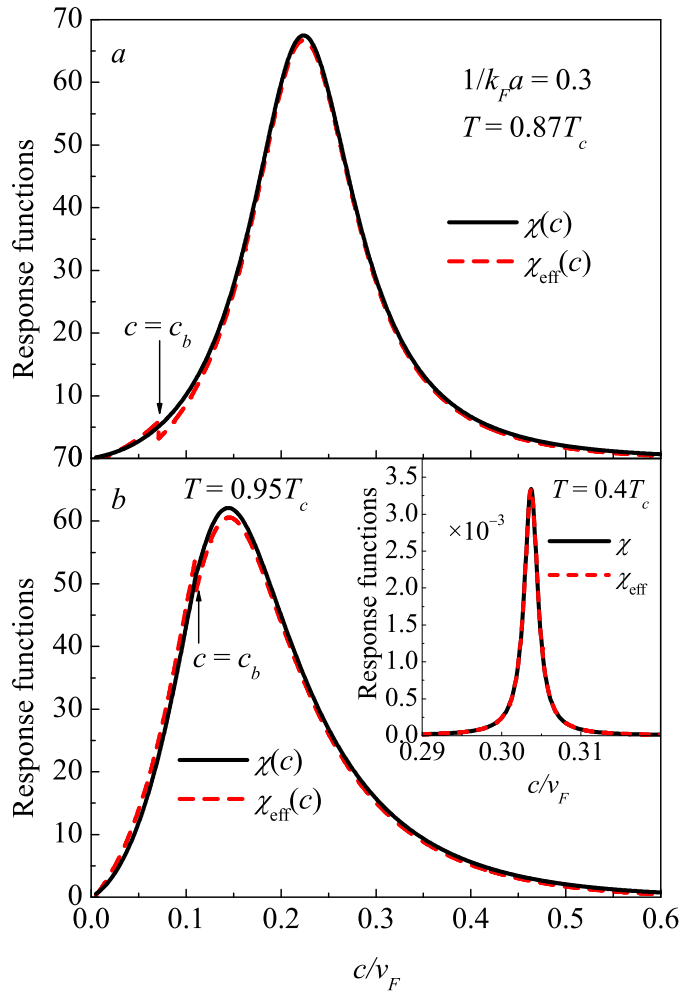}%
\caption{(Color online) Long-wavelength phase-phase response function
$\chi\left(  c\right)  $ and its two-pole analytic approximation
$\chi_{\mathrm{eff}}\left(  c\right)  $ at $1/k_{F}a=0.3$). At $T=0.4T_{c}$
(inset), both show a single quasi-Lorentzian peak. At $T=0.87T_{c}$ (panel
$a$), the boundary velocity $c_{b}$, which is an angular point for $\chi$ and
a discontinuity for $\chi_{\mathrm{eff}}$, lies below the maximum of the
resonance. At $T=0.95T_{c}$ (panel $b$) the angular point shifts towards
higher $c$.}%
\label{fig:rep2}%
\end{figure}

On the contrary, when $c_{\mathrm{s}1}$ or $c_{\mathrm{s}2}$ cross $c_b$
at a given temperature (which occurs with the scaled GPF equation of state
for $0.679\gtrapprox1/k_Fa\gtrapprox-0.594$;
in Fig.~\ref{fig:svUn} we show the example of unitary $1/|a|=0$),
this means that the angular point in $c=c_b$ goes through the peak of $\chi$ as temperature varies,
as illustrated in Fig.~\ref{fig:rep2}. Then, the roots found in window $A$ of the analytic continuation
describe the left part of this broken peak, and those of window $B$, its right part.
In practice, when they are close to $c_{b}$, the difference between the sound velocities $c_{\text{s}%
}^{\left(  A\right)  }$ and $c_{\text{s}}^{\left(  B\right)  }$ in the two windows is small
with respect to their imaginary parts $\kappa_{\text{s}}$, as can be seen from Fig.~\ref{fig:svUn}.
Physically, since the damping factor is a measure of the uncertainty of the sound
velocity (following from the uncertainty relation between time and energy),
this means that the difference in the velocity is almost indistinguishable, or
in other words, that the discontinuity in the slope of the resonance peak
can only be resolved through a very precise measurement of the response function.

\paragraph{Analytic approximation for the response function}
\label{sec:fit}

From the poles $u_{\textrm{s},1}$ and $u_{\textrm{s},2}$ found in the analytic continuation,
and their residues $Z_1$ and $Z_2$ in the phase-phase propagator $\Im m_{--}/\pi W$
one can construct an effective response function, in the BCS regime:
\begin{equation}
\label{chieff1}
\chi_{\mathrm{eff}}\left(  c\right)  = \begin{cases}
\frac{1}{\pi}\operatorname{Im}\left(
\frac{Z_{1}^{(A)}}{c-u_{\mathrm{s1}}^{(A)}}+\frac{Z_{2}^{(A)}}{c-u_{\mathrm{s2}}^{(A)}}\right) \quad\textrm{if}\quad 0\leqslant c \leqslant c_b \\
\frac{1}{\pi}\operatorname{Im}\left(
\frac{Z_{1}^{(B)}}{c-u_{\mathrm{s1}}^{(B)}}+\frac{Z_{2}^{(B)}}{c-u_{\mathrm{s2}}^{(B)}}\right) \quad\textrm{if}\quad c\geqslant c_b,
\end{cases}
\end{equation}
which is the sum of the two resonance peaks caused by
$u_{\textrm{s},1}$ and $u_{\textrm{s},2}$ in each window $A$ and $B$.
Note that since the residues $Z_1$ and $Z_2$ are complex, this is not
simply the sum of two Lorentzian functions. Conversely, in the BEC regime,
our effective response function has only one resonance
\begin{equation}
\label{chieff2}
\chi_{\mathrm{eff,B}}\left(  c\right)  =
\frac{1}{\pi}\operatorname{Im}\left(
\frac{Z_{\rm B}}{c-u_{\mathrm{s,B}}}\right)
\end{equation}
These functions can be compared with the exact response function $\chi$,
to check the relevance of the analytic structure found in the analytic continuation.
They allow to interpret the shape of $\chi$ in terms of resonances caused by collective modes.
They can also be used as fitting functions for experimentalists
to extract the values of $u_{\textrm{s},1}$,
$u_{\textrm{s},2}$ or $u_{\textrm{s,B}}$ and their residues from
a measured response spectrum.


In the low-temperature case (see the example of $T=0.4T_{c}$ in the inset
$(a)$ of Fig.~\ref{fig:rep2}) the residue of the only relevant complex root
tends to a real number, such that we expect the response function $\chi$ to
have an approximate Lorentzian shape. This is indeed what we observe in
Fig.~\ref{fig:rep2}, with a very good agreement between $\chi$ and
$\chi_{\mathrm{eff}}$. When raising the temperature,
away from the BCS regime one does not immediately observe the
formation of a second peak (see the examples of
$T=0.87T_{c}$ and $T=0.95T_{c}$ in Fig.~ \ref{fig:rep2}), but rather a shift
in the position of the original peak and an increase of its width and
skewness. To describe the altered peak, we introduce an effective sound
velocity $u_{\mathrm{s,eff}}=c_{\mathrm{s,eff}}-i\kappa_{\mathrm{s,eff}}$
where $c_{\mathrm{s,eff}}$ is the value of $c$ where $\chi_{\mathrm{eff}}$
reaches its maximum, and $\kappa_{\mathrm{s,eff}}$ its half width at half
maximum\footnote{The effective sound velocity $c_{\mathrm{s,eff}}$ is given
analytically by the equation%
\begin{equation}
\operatorname{Im}\left(  \frac{Z_{1}}{\left(  c_{\mathrm{s,eff}}%
-u_{\mathrm{s1}}\right)  ^{2}}+\frac{Z_{2}}{\left(  c_{\mathrm{s,eff}%
}-u_{\mathrm{s2}}\right)  ^{2}}\right)  =0. \label{equ1}%
\end{equation}
The effective damping factor
is the half width of $\chi_{\mathrm{eff}}$ at its
half height:%
\begin{equation}
\kappa_{\mathrm{s,eff}}=\frac{1}{2}\left(  c_{\mathrm{hw}}^{\left(  2\right)
}-c_{\mathrm{hw}}^{\left(  1\right)  }\right)  , \label{happa1}%
\end{equation}
where $c_{\mathrm{hw}}^{\left(  1\right)  }<c_{\mathrm{s,eff}}$ and
$c_{\mathrm{hw}}^{\left(  2\right)  }>c_{\mathrm{s,eff}}$ are the two roots
of the equation:%
\begin{equation}
\chi_{\rm eff}\left(  c_{\mathrm{hw}}\right)  =\frac{1}{2}\chi_{\rm eff}\left(  c_{\mathrm{s,eff}%
}\right)  . \label{hw}%
\end{equation}
Naturally, these definitions are valid only when $\chi_{\rm eff}$ shows a single maximum.}.
These quantities are useful only close to $T=0$ and $T_c$ where one root is much
less damped than the other.
In the intermediate temperature regime where the two roots have a comparable damping rate,
the response function is not well fitted by a single Lorentzian,
and one should revert to the superposition introduced in \eqref{chieff1}.
This particularly the case in the far BCS regime where the response function
exhibits two distinct peaks in a temperature range close to but excluding $T_c$
(see the example of $T=0.95T_{c}$ in Fig.~\ref{fig:rep}).

As we said above, at and around unitarity, the angular point in $c_b$
goes through the resonance peak as temperature varies.
This results in a visibly broken peak in $\chi$, which is again
well captured by our two-pole analytic approximation $\chi_{\rm eff}$ provided one switches of the interval
of analytic continuation when crossing $c_b$, as prescribed by Eq.~\eqref{chieff1}.
When the argument $c=\omega/q$ of the response function passes the boundary velocity $c_b$, $\chi\left(  c\right)$
exhibits an angular point and its two-pole analytic approximation
$\chi_{\mathrm{eff}}\left(  c\right)$ exhibits a discontinuity.

\section{Links to other theories and to experiments \label{sec:links}}

\subsection{Comparison to low temperature approaches}%

\begin{figure}[tbh]%
\centering
\includegraphics[
width=3.5in
]%
{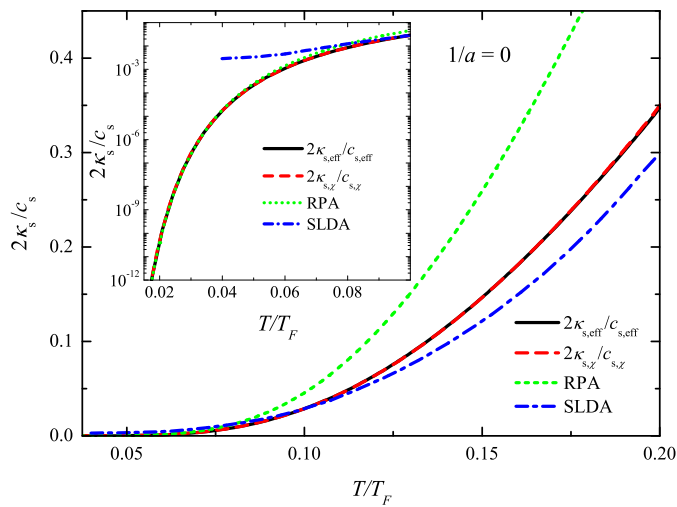}%
\caption{(Color online) Long-wavelength inverse quality
factor of the
Bogoliubov-Anderson mode at unitarity and at low temperature.
\textit{Solid curve}: An effective quality factor is computed from
the poles $u_{s1}$ and $u_{s2}$ in the analytic continuation
by taking into account the unicity of the resonance at low temperature
(see Eqs.~\eqref{equ1} and \eqref{happa1}). \textit{Dashed curve}:
The quality factor is extracted directly from
the long-wavelength response function $\chi\left(  c\right)  $.
In both cases we use the equation of state obtained within the GPF approximation \cite{HLD},
instead of the mean-field one.
\textit{Dotted curve}: the RPA low-temperature asymptotic behavior
according to \cite{Kurkjian2017-2}
is recalculated using the GPF equation of state. \textit{Dotted
curve}: the SLDA result of Ref. \cite{Zou}. \textit{Inset}: the same in a
lower temperature range,  in the logarithmic scale.}%
\label{fig:damp}%
\end{figure}
In Fig. \ref{fig:damp}, we plot the inverse quality factor $2\kappa
_{\mathrm{s,eff}}/c_{\mathrm{s,eff}}$ of the phononic modes as a function of the
temperature at unitarity where we use the scaled GPF equation of state.
In this regime, our result can be compared to several other approaches (which all assumed
the existence of a unique phononic mode, hence our use of the effective velocity).
(i) A prediction based on Landau phonon-roton theory
(which is exact if the roton branch is known exactly, see
Eqs.~(15-16) of \cite{Kurkjian2017-3}; it is recalculated here using the BCS
branch as the roton branch and the GPF parameters of state\footnote{This RPA
result is erroneously reproduced in Ref. \cite{Zou} because an incorrect value
for the parameter $d\Delta/d\mu=-0.58$ was used. The correct parameter at
unitarity is $\left.  d\Delta/d\mu\right\vert _{T=0}=\left.  \Delta
/\mu\right\vert _{T=0}\approx1.162$. This gives us the pre-exponential factor
in the low-temperature expansion of $\lim_{q\rightarrow0}\left(  \Gamma
_{q}/\omega_{q}\right)  $ approximately equal to 8.37 instead of the value 1.6
used in Ref. \cite{Zou}.}) exactly agrees with our asymptotic results when\footnote{
The damping of phononic modes in Ref. \cite{Kurkjian2017-2}
has been calculated within the perturbative approach, which is valid for sufficiently
low temperatures, but exhibits a difference with the present non-perturbative
method for $k_{B}T/E_{F}\gtrsim0.1$.
}
$T\rightarrow0$.
(ii) The superfluid local density approximation (SLDA)
\cite{Zou}, an approach which exploits the universal behavior of the gas at
unitarity, also predicts a quality factor due to the coupling to the fermionic
quasiparticle-quasihole continuum in good quantitative agreement with ours
except the low-temperature range where the damping rate obtained in \cite{Zou}
seems aberrant as it does not tend to zero when $T\rightarrow0$.

\subsection{Comparison to measurements of the sound velocity}

\label{sec:comparison}

\begin{figure}[hbt]%
\centering
\includegraphics[
width=3.5in
]%
{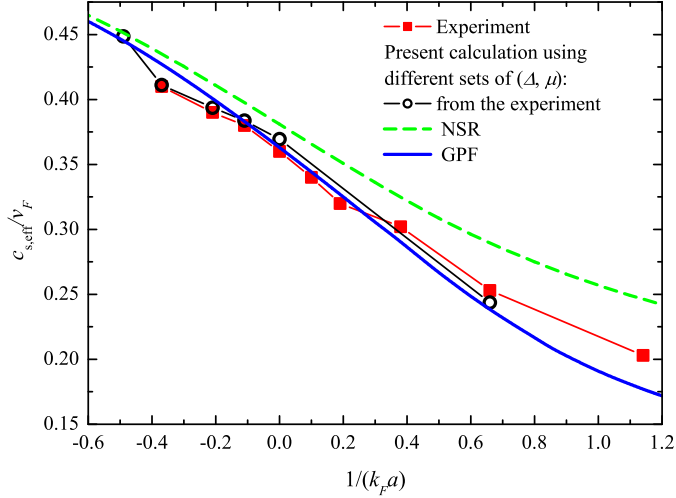}%
\caption{(Color online) Nonzero-temperature sound velocity $c_{\mathrm{s}}$ as
a function of $1/k_Fa$ calculated within the present approach using different
sets of background parameters $\left(  \mu,\Delta\right)  $: with background
parameters extracted from \cite{Hoinka} (\textit{empty dots}), with
parameters calculated accounting for Gaussian fluctuations within the NSR
scheme for the superfluid (broken-symmetry) state \cite{Engelbrecht}
(\textit{dashed curve}) and
within the GPF scheme of Refs. \cite{Diener2008,HLD,HLD1} (\textit{solid curve}). The
calculated sound velocities are compared with the experimental data of Ref.
\cite{Hoinka} (\textit{squares}).}%
\label{fig:exp}%
\end{figure}

In Fig. \ref{fig:exp}, the nonzero-temperature effective sound velocity
$c_{\mathrm{s,eff}}$ as a function of $1/k_Fa$ calculated within the present
approach is compared with the experimental data of Ref. \cite{Hoinka}
(squares) using different equations of state. Since the experimental value of the speed of sound
were obtained using a single Gaussian fit of the response function,
it is natural to compare them to our effective sound velocity (which combines
the information about the two resonances in a unique velocity).

The temperatures throughout the BCS-BEC crossover are
determined by a quadratic interpolation of the experimental values reported in
Ref. \cite{Hoinka}: $k_{B}T=0.09E_{F}$ at unitarity, $k_{B}T=0.02E_{F}$ at
$1/k_{F}a=-1.6$, and $k_{B}T=0.1E_{F}$ at $1/k_{F}a=1$ (such that $T/T_c$ is about $1/2$
in all three cases). The sound velocity has
been calculated here using our results for $c_s(\Delta/\mu,\Delta/T)$,
and the mean-field gap equation with the chemical
potential obtained by three methods: (1) from the Table
3 of the Supplement to Ref. \cite{Hoinka}, (2) from the number equation accounting for
Gaussian pair fluctuations within the NSR scheme for the superfluid state
below $T_{c}$ \cite{Engelbrecht}, and (3) from the GPF approach of Refs.~\cite{Diener2008,HLD,HLD1}
(almost equivalent to the scaled GPF equation of state of Appendix \ref{app:eos} since the temperature is lower than $T_c$ here).
As can be seen from Fig. \ref{fig:exp}, an excellent agreement with the experiment is obtained when
we use the parameters of state from Ref. \cite{Hoinka}.

\subsection{Measuring the phase-phase response}

\begin{figure}[h!]
\begin{center}
\begin{tikzpicture}
\tikzstyle{corde}=[fill,pattern=dots,minimum width=0.75cm,minimum height=0.3cm]
\tikzstyle{mur}=[fill,pattern=north east lines,draw=none,minimum width=0.75cm,minimum height=0.3cm]
\tikzset{paire/.pic={\draw[rotate=-45,dashed] (0,0) ellipse (0.4 and 0.15);
				\draw[red] (-0.2,0.2) node {\tiny{\textbullet}};
				\draw[blue] (0.2,-0.2) node {\tiny{\textbullet}};}}
\tikzset{paire2/.pic={\draw[rotate=45,dashed] (0,0) ellipse (0.4 and 0.15);
				\draw[red] (0.2,0.2) node {\tiny{\textbullet}};
				\draw[blue] (-0.2,-0.2) node {\tiny{\textbullet}};}}
\tikzset{paire3/.pic={\draw[dashed] (0,0) ellipse (0.4 and 0.15);
				\draw[red] (0.3,0) node {\tiny{\textbullet}};
				\draw[blue] (-0.3,0) node {\tiny{\textbullet}};}}
\tikzset{paire4/.pic={\draw[rotate=-20,dashed] (0,0) ellipse (0.4 and 0.15);
				\draw[rotate=-20,red] (0.3,0) node {\tiny{\textbullet}};
				\draw[rotate=-20,blue] (-0.3,0) node {\tiny{\textbullet}};}}
\tikzset{paire5/.pic={\draw[rotate=20,dashed] (0,0) ellipse (0.4 and 0.15);
				\draw[rotate=20,red] (0.3,0) node {\tiny{\textbullet}};
				\draw[rotate=20,blue] (-0.3,0) node {\tiny{\textbullet}};}}
\draw[->] (-4,-2.1) -- (-4,0) node {$-$} node[left] {$0$} -- (-4,2.5) node[left] {$y$};
\draw (0,-2) node[below] {$L$} -- (-4,-2) node[below left] {$0$} -- (-4,2) -- (0,2);
\draw [domain=-2:2,samples=200] plot ({0.1*cos(3.14*(6+1)*\x r/4)},\x);
\draw (4,-2) -- (0.5,-2)  (4,2) -- (0.5,2);
\draw[dashed] (6,2) -- (4,2) ;
\draw[dashed,->] (4,-2) -- (6,-2) node[below] {$x$};
\draw [domain=-2:2,samples=200] plot ({0.5-0.1*cos(3.14*(6+1)*\x r/4)},\x);
\draw (-3,-1) pic{paire};
\draw (-2.5,1) pic{paire2};
\draw (-0.8,0.3) pic{paire4};

\draw [red,domain=-3.9:-0.3,only marks, mark=*, samples=20, mark size=1.] plot (\x,{0+1.9*rand});
\draw [blue,domain=-3.9:-0.3,only marks, mark=*, samples=20, mark size=1.] plot (\x,{0+1.9*rand});

\draw (3,0) pic{paire4};
\draw (3,1) pic{paire2};
\draw (3.5,0.5) pic{paire};
\draw (2.5,1) pic{paire4};
\draw (2.5,1.7) pic{paire3};
\draw (1.5,0) pic{paire};
\draw (1,1) pic{paire2};
\draw (3.6,-1.7) pic{paire};
\draw (1.2,-1.1) pic{paire5};
\draw (2.6,-1.2) pic{paire4};
\draw (2.4,-0.2) pic{paire3};
\draw (1.8,0.8) pic{paire5};
\draw (1.2,-1.8) pic{paire4};
\draw (2.5,-1.8) node {${\large T=0}$};
\draw (-2,-1.8) node {${\large T\to T_c}$};
\end{tikzpicture}
\end{center}
\caption{\label{fig:mesure} Experimental setup designed to measure the phase-phase response function $\chi$ of a superfluid Fermi gas at $T\lessapprox T_c$ (left well). A spatially dependent tunneling barrier (in the middle) couples the system to a reservoir (right well) consisting of a fully paired Fermi gas at $T=0$. This creates a spatially dependent excitation of the phase of the order parameter in the left well. This phase pattern is measured at the end of the experimental sequence by letting the gases in the left and right wells expand and interfere.}
\end{figure}
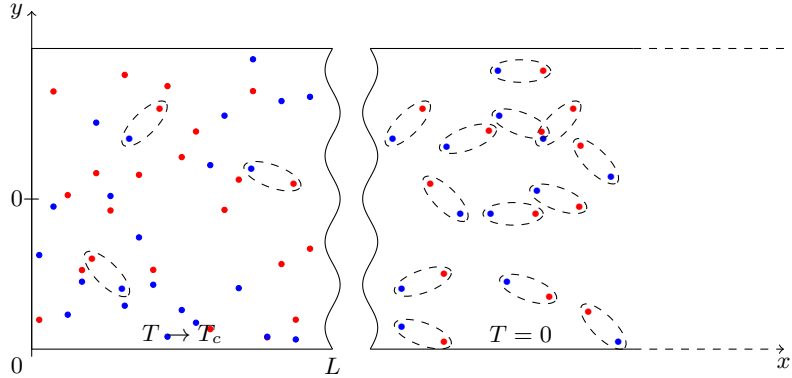

So far the experiments have measured the collective mode
spectrum through the density response of the gas. It would be interesting
to access also the phase-phase response function, particularly near $T_c$
where it has a very different shape than the density-density response as we have seen.
To this end, we explain how one can adapt the Carlson-Goldman
\cite{Goldman1976} experiment, which measured the pairing-field
susceptibility of a superconductor, to a cold atom setup.
The scheme we propose is illustrated on Fig.~\ref{fig:mesure},
and it uses only existing experimental techniques.

The excitation is obtained by coupling
the system of interest (a superfluid Fermi gas at nonzero temperature, for example at
$T$ close to $T_c$) to a environment consisting of a large superfluid Fermi gas
prepared at zero temperature and with a well-defined phase with respect to the system (which can be done
by initially performing Josephson oscillations \cite{Roati2015}). The two gases are coupled through a tunneling barrier,
similar to the thin barrier realized in \cite{Roati2015}; to extract information on the spectrum at momentum $q$,
the barrier should be spatially modulated at a wavelength $\lambda=2\pi/q$, which, for instance, could be achieved
by interfering two laser fields.
The fact that the reservoir gas is much larger than the studied system ensures that it remains at zero temperature
all along the excitation time, and that its quantum fluctuations can be neglected. It behaves then
as a classical pairing field imposed on the system, which can be represented by the drive term in the Hamiltonian
\be
\hat H_{\rm drive}(t)=\int d^3 r J(\textbf{r},t) \Delta_{\rm exc}^* \hat\psi_\downarrow(\textbf{r},t) \hat\psi_\uparrow(\textbf{r},t) + \mbox{h.c.}
\ee
Here $\Delta_{\rm exc}$ is the order parameter of the reservoir (whose phase has been fixed initially) and $J(\textbf{r},t)=J(t)\times\cos(qy)$ is the spatially
dependent strength of the barrier; since the system is prepared close to $T_c$, its healing length is very large, and we can
assume the effect of the barrier to be homogeneous in the $x$-direction \cite{Goldman1976}. The time-dependence of $J$
can be either sinusoidal $J(t)\propto \cos(\omega t + \phi)$ if one wishes to probe the response function at a given frequency $\omega$,
or it can be more abrupt if one wishes to study the quench-like dynamics of the system (which theoretically is described by the Laplace transform
of the frequency-domain response function $\chi$ \cite{Gurarie2009}). Finally, the phase of the system is measured by letting the cloud expand and interfere with the reservoir as in \cite{Roati2015}. The interference pattern will appear shifted in the $x$ direction by a length $\delta x(y)$ which depends on the local phase of the system at position $y$.

\section{Conclusions \label{sec:conclusions}}

We have studied the long-wavelength solutions to the RPA/GPF
equation on the collective mode energy of a neutral fermionic condensate.
To access the full range of temperatures
between zero and $T_c$, we deal non-perturbatively with the damping caused
by absorption/emission of BCS ``broken-pair'' quasiparticles.
For that, we set the energy proportional to the wave vector, $z=uq$,
and analytically continue the equation for $u$ through its branch cut
associated to the quasiparticle absorption-emission continuum.

While our results at low temperature agree with previous perturbative
approaches in predicting a single collective mode with
an exponentially small damping rate and velocity shift,
we find an unexpected second solution in the vicinity of the transition temperature $T_c$.
This two-mode nature is also visible in the order-parameter phase response function
which displays two distinct resonance peaks, at temperatures relatively close to $T_c$,
and in the BCS regime.
In the limit $T\to T_c$, we show analytically that the velocity of the first mode tends to a finite
and non-zero complex number, while the damping rate of the second mode vanishes like $\Delta(T)$ (or $(T_c-T)^{1/2}$),
and its quality factor vanishes logarithmically. In the BEC regime, on the contrary, we find only one
relevant solution, whose velocity vanishes like $(T_c-T)^{1/2}$ near $T_c$ with a diverging quality factor.

At arbitrary temperatures $0<T<T_c$, we develop a numerical method
to perform the analytic continuation of the GPF equation. This confirms
the existence of two distinct phononic branches, one being dominant near $T=0$,
the other near $T_c$. The transition between these two resonances is visible in both
the phase-phase and density-density response functions.
Last, our knowledge of the two poles in the analytic
continuation, and of their residues, allows us to propose an analytic function
to describe the phase response in terms of two collective resonances.

The present study not only resolves some problems but also raises new questions,
particularly about the existence, outside the collisionless regime, of the transition we have seen between
two distinct collective modes. This transition undoubtedly exists in the GPF approximation,
but a more systematic treatment should account for the finite lifetime of the
fermionic quasiparticles \cite{Zwerger2009}.
In any case, our work will be heuristically useful for further developments
of the theory of collective excitations in superfluid Fermi gases.

\begin{acknowledgments}
We thank C. A. R. S\'{a} de Melo and M. W. Zwierlein for valuable discussions.
This research was supported by the University Research Fund (BOF) of the
University of Antwerp and by the Flemish Research Foundation (FWO-Vl), project
No. G.0429.15.N. and the European Union's Horizon 2020 research and innovation
program under the Marie Sk\l odowska-Curie grant agreement number 665501.
\end{acknowledgments}

\appendix

\section{Equation of state accounting for order-parameter fluctuations}
\label{app:eos}

The next-order approximation beyond mean-field is used to account for fluctuations
of the pair field about the saddle-point solution. This comes from the expansion of
the effective bosonic action in powers of the fluctuation coordinates up to
the second order. The resulting thermodynamic potential is a sum of the
saddle-point and fluctuation contributions:%
\begin{equation}
\Omega=\Omega_{\rm sp}+\Omega_{\rm fluct},
\end{equation}
with the saddle-point and fluctuation contributions:%
\begin{subequations}
\begin{align}
\Omega_{\rm sp} &  =-\int\frac{d^{3}k}{\left(  2\pi\right)  ^{3}}\left[  \frac
{1}{\beta}\ln\left(  2+2\cosh\beta E_{\mathbf{k}}\right)  -\xi_{\mathbf{k}%
}-\frac{m\Delta^{2}}{k^{2}}\right]  -\frac{m\Delta^{2}}{4\pi a},\\
\Omega_{\rm fluct} &  =\frac{1}{2\beta}\sum_{\mathbf{q},n}\ln\det\mathbb{M}\left(
\mathbf{q},i\Omega_{n}\right)  ,
\end{align}
where the matrix elements of the inverse Gaussian pair fluctuation $M\left(
\mathbf{q},i\Omega_{n}\right)  $ propagator are described above.

Within the Nozi\`{e}res -- Schmitt-Rink (NSR) scheme \cite{deMelo1993}
extended to the superfluid state below $T_{c}$ in Ref. \cite{Engelbrecht} (see
also \cite{Ohashi2006,PRA2008}), the particle density is determined
as%
\end{subequations}
\begin{equation}
n=-\left.  \left(  \frac{\partial\Omega}{\partial\mu}\right)  \right\vert
_{T,\Delta}\label{nnsr}%
\end{equation}
considering $\Delta$ as an independent parameter. The NSR scheme has been
modified \cite{Diener2008,HLD} taking into account a variation
of the gap:%
\begin{equation}
n=-\left.  \left(  \frac{\partial\Omega}{\partial\mu}\right)  \right\vert
_{T,\Delta}-\left.  \left(  \frac{\partial\Omega}{\partial\Delta}\right)
\right\vert _{T,\mu}\left.  \left(  \frac{\partial\Delta}{\partial\mu}\right)
\right\vert _{T}.\label{ngpf}%
\end{equation}
This approximation, referred to as GPF (Gaussian Pair Fluctuation
approximation) provides the temperature dependence of the chemical potential
in good agreement with Quantum Monte Carlo results \cite{Astr,Bulgac2006}.

The parameters of state accounting for fluctuations are related to the
saddle-point parameters of state through the exact scaling relations%

\begin{equation}
\Delta\left(  \frac{1}{a},T,\mu\right)  =\Delta_{sp}\left(  \frac{1}{a_{sp}%
},T_{sp},\mu_{sp}\right)  \frac{E_{F}^{\left(  sp\right)  }\left(  \frac
{1}{a_{sp}},T_{sp},\mu_{sp}\right)  }{E_{F}\left(  \frac{1}{a_{sp}},T_{sp}%
,\mu_{sp}\right)  }\label{sc1}%
\end{equation}
where $\left(  \frac{1}{a_{\rm sp}},T_{\rm sp},\mu_{\rm sp}\right)  $ are related to the
true $\left(  \frac{1}{a},T,\mu\right)  $ by the equations:%
\begin{align}
\frac{a}{a_{\rm sp}}  & =\sqrt{\frac{E_{F}\left(  \frac{1}{a_{\rm sp}},T_{\rm sp}\right)
}{E_{F}^{\rm \left(  sp\right)  }\left(  \frac{1}{a_{\rm sp}},T_{\rm sp}\right)  }%
},\label{sc2}\\
\frac{T}{T_{\rm sp}}  & =\frac{\mu}{\mu_{\rm sp}}=\frac{E_{F}^{\rm \left(  sp\right)
}\left(  \frac{1}{a_{\rm sp}},T_{\rm sp}\right)  }{E_{F}\left(  \frac{1}{a_{\rm sp}%
},T_{\rm sp}\right)  },\label{sc3}%
\end{align}
with the Fermi energies $E_{F}$ and $E_{F}^{\left(  sp\right)  }$
calculated, respectively, with and without accounting for fluctuations:%
\begin{equation}
E_{F}=\frac{\hbar^{2}\left(  3\pi^{2}n\right)  ^{2/3}}{2m},\quad
E_{F}^{\left(  sp\right)  }=\frac{\hbar^{2}\left(  3\pi^{2}n_{sp}\right)
^{2/3}}{2m}.
\end{equation}
These scaling relations precisely reproduce the GPF or NSR schemes [depending
of a choice for $n$, (\ref{nnsr}) or (\ref{ngpf})]. Close to the transition
temperature, both NSR and GPF schemes reveal an artifact: a discontinuous
change of the gap from a finite value to zero at $T_{c}$. In order to overcome
this issue and study sound velocities in a superfluid Fermi gas for all
$T<T_{c}$, several interpolation schemes were considered in Refs.
\cite{Taylor2008,Taylor}. In the present work, we use a slightly different
interpolation scheme. The chemical potential calculated within the GPF
approach \cite{HLD,HLD1} shows an excellent agreement with the Monte Carlo
results for $T<T_{c}$ \cite{Bulgac2006}, where the transition temperature
$T_{c}$ is determined accounting for fluctuations and is the same within the
GPF and NSR schemes \cite{HLD1}. Moreover, both the chemical potential and the
gap calculated within GPF at $T=0$ are in good agreement with these Monte
Carlo calculations. Therefore we keep the relations (\ref{sc2}) and
(\ref{sc3}) unchanged, so that the chemical potential remains the same as
within GPF, and replace (\ref{sc1}) by the equation:%
\begin{equation}
\Delta\left(  \frac{1}{a},T,\mu\right)  =\Delta_{sp}\left(  \frac{1}{a_{sp}%
},T_{sp}^{\prime},\mu_{sp}\right)  \frac{E_{F}^{\left(  sp\right)  }\left(
\frac{1}{a_{sp}},T_{sp},\mu_{sp}\right)  }{E_{F}\left(  \frac{1}{a_{sp}%
},T_{sp},\mu_{sp}\right)  },\label{sc1a}%
\end{equation}
where $T_{\rm sp}$
in the temperature dependence of $\Delta_{\rm sp}$
is rescaled as $T_{\rm sp}^{\prime}%
\equiv\left(  T_{c}^{\rm \left(  sp\right)  }/T_{c}\right)  T$. According to
(\ref{sc1a}), the gap takes the value $\Delta=\Delta_{\rm GPF}$ at $T=0$, and
tends to zero as $\Delta\propto\sqrt{T_{c}-T}$ when approaching $T_{c}$. This
known behavior of $\Delta$ in the vicinity of $T_{c}$ is an exact universal
condition, independent on the coupling strength. Eq. (\ref{sc1a}) is thus a
renormalized saddle-point gap equation in which the aforesaid artifacts of the
temperature dependence of $\Delta\left(  T\right)  $ are removed.

\section{Details of the calculation near $T_c$}
\label{app:Tc}
In this Appendix, we detail the calculation of the integrals in Eqs.~(\ref{mpp}--\ref{mpm}) in the limit $T\to T_c$. We recall the notations of Sec.~\ref{sec:Tc}: $\epsilon=\Delta/T$ is our small parameter, $\mu/T=m_c+O(\epsilon^2)$, and $\mu/\Delta=m_c/\epsilon+O(\epsilon)$. We first replace the tridimensional integral in the following way
\be
 \int \frac{d^3 k}{(2\pi)^3} \to \frac{\rho(\mu)\Delta}{2}\int_{-m_c/\epsilon}^\infty \check{k}(\xi) d\xi \int_0^1 dt \label{integraleapp}
\ee
where $\rho(\mu)=\sqrt{2m^3\mu}/\pi^2\hbar^3$ is the density of states at energy $\mu$ (setting the gas volume equal to $1$), $t=\cos\theta$, $\xi=\xi_k/\Delta$, $\check{k}^2=k^2/2m\mu=1+\epsilon\xi/m_c+O(\epsilon^3)$ and we use later $E^2=\xi^2+1$.

\paragraph{BCS regime}
In the BCS regime ($m_c>0$), we give the formulary of $\epsilon$-expanded integrals:
\bea
\int_{-m_c/\epsilon}^\infinity d\xi\frac{\check{k}^p(\xi)\textrm{tanh}(\epsilon E/2)}{E^3} &=& \epsilon\frac{\pi}{2} + \epsilon^2 f_p(m_c)+O(\epsilon^3), \qquad p=1,3 \label{formulaire1}\\
\int_{-m_c/\epsilon}^\infinity d\xi\frac{\check{k}^p(\xi)\textrm{tanh}'(\epsilon E/2)}{E^2} &=&  {\pi}+\epsilon g_p(m_c) O(\epsilon^2), \qquad p=1,3 \\
\int_{-m_c/\epsilon}^\infinity d\xi\frac{\check{k}(\xi)\textrm{tanh}(\epsilon E/2)}{E^3} \xi &=& \epsilon h_1(m_c)+O(\epsilon^3)\label{formulaire3}
\eea
The leading orders are obtained by simply expanding the integrand at low $\epsilon$: ${\check{k}^p(\xi)\textrm{tanh}(\epsilon E/2)}/{E^n}=\epsilon /{2E^{n-1}}+O(\epsilon^2)$, and ${\check{k}^p(\xi)\textrm{tanh}'(\epsilon E/2)}/{E^n}=1 /{E^n}+O(\epsilon^2)$ which gives rise to converging integrals. To compute the subleading term, we add and subtract the leading one to ensure the convergence of the $\epsilon$-expanded integral in $\xi=0$, perform the change of variable $e=\epsilon\xi$, and approximate $\epsilon E\simeq e$. We get
\begin{multline}
f_p(m_c)=-\frac{1}{m_c} + \int_{0}^{m_c} de\bbaco{\frac{\textrm{tanh}(e/2)}{e^3}\bbcro{\bb{1+\frac{e}{m_c}}^{p/2}+\bb{1-\frac{e}{m_c}}^{p/2}}-\frac{1}{e^2}} \\
+\int_{m_c}^\infty de \frac{\textrm{tanh}(e/2)}{e^3} \bb{1+\frac{e}{m_c}}^{p/2} \qquad p=1,3 \label{grosseint1}
\end{multline}
\begin{multline}
g_p(m_c)=-\frac{2}{m_c} + \int_{0}^{m_c} de\bbaco{\frac{\textrm{tanh}'(e/2)}{e^2}\bbcro{\bb{1+\frac{e}{m_c}}^{p/2}+\bb{1-\frac{e}{m_c}}^{p/2}}-\frac{2}{e^2}} \\
+\int_{m_c}^\infty de \frac{\textrm{tanh}'(e/2)}{e^2} \bb{1+\frac{e}{m_c}}^{p/2} \qquad p=1,3 \label{grosseint2}
\end{multline}
\be
h_1(m_c)= \int_{0}^{m_c} de\frac{\textrm{tanh}(e/2)}{e^2}\bbcro{\bb{1+\frac{e}{m_c}}^{1/2}-\bb{1-\frac{e}{m_c}}^{1/2}}
+\int_{m_c}^\infty de \frac{\textrm{tanh}(e/2)}{e^2} \bb{1+\frac{e}{m_c}}^{1/2} \label{grosseint3}
\ee
In $m_{++}$, $m_{--}$, $m_{+-}$, the integrals with a resonant denominator give
\bea
\int_{-m_c/\epsilon}^\infinity d\xi \int_0^1 dt
\frac{ \check{k}^3t^2 \textrm{tanh}'\bb{{\epsilon E}/{2}}}{E^2(E^2\check{u}^2-t^2\xi^2\check{k}^2)} &=&  \pi + 4F(\check{u})+O(\epsilon^2) \label{intu1}\\
\int_{-m_c/\epsilon}^\infinity d\xi \int_0^1 dt
\frac{ \check{k}^3t^2 \textrm{tanh}'\bb{{\epsilon E}/{2}}\xi^2}{E^2(E^2\check{u}^2-t^2\xi^2\check{k}^2)} &=&  -\pi + 4G(\check{u})-\epsilon G_2(\check{u},m_c) + O(\epsilon^2)
 \\
\int_{-m_c/\epsilon}^\infinity d\xi \int_0^1 dt
\frac{ \check{k}^3t^2 \textrm{tanh}'\bb{{\epsilon E}/{2}}\xi}{E^2(E^2\check{u}^2-t^2\xi^2\check{k}^2)} &=& \epsilon H(\check{u},m_c)+O(\epsilon^2) \label{intu3}
\eea
The functions $F$ and $G$ are given in the main text [Eqs.~\eqref{F}-\eqref{G}]. Functions $G_2$ and $H$ characterizing the subleading order terms to $m_{--}$ and $m_{+-}$ can be written in integral forms similar to Eqs.~(\ref{grosseint1}--\ref{grosseint3}), which we don't give explicitly. In fact, we will need only the value of these functions in $\check{u}$=0:
\bea
G_2(0,m_c)&=& g_1(m_c) \\
H(0,m_c)&=& 0
\eea
Combining our two formularies (\ref{formulaire1}--\ref{formulaire3}) and (\ref{intu1}--\ref{intu3}), to the definition of $m_{\sigma\sigma'}$ (Eqs.~(\ref{mpp}--\ref{mpm})), we obtain equations (\ref{mppTc}--\ref{mpmTc}) of the main text, with
\bea
f&=&\frac{f_3}{6} -\frac{g_3}{12}\\
g&=&\frac{f_1}{2}-\frac{g_1}{4}\\
h&=&-\frac{h_1}{4}
\eea

\paragraph{BEC regime}
In the BEC regime ($m_c<0$), the integral over $\xi$ in \eqref{integraleapp} begins from $|{m_c}|/\epsilon\gg1$.
To leading order, one then approximates $E\simeq\xi$, and performs the change of variable $e=\epsilon\xi$. We give the new formulary of integrals
\bea
\int_{|m_c|/\epsilon}^\infinity d\xi\frac{\check{k}^p(\xi)\textrm{tanh}(\epsilon E/2)}{E^3} &=& \epsilon^2 f_p^{\rm B}(m_c)+O(\epsilon^4), \qquad p=1,3 \label{formulaireBEC1}\\
\int_{|m_c|/\epsilon}^\infinity d\xi\frac{\check{k}^p(\xi)\textrm{tanh}'(\epsilon E/2)}{E^2} &=&  \epsilon g_p^{\rm B}(m_c) + O(\epsilon^3), \qquad p=1,3 \\
\int_{|\mu|/\Delta}^\infinity d\xi\frac{\check{k}(\xi)\textrm{tanh}(\epsilon E/2)}{E^3} \xi &=& \epsilon h_1^{\rm B}(m_c)+\epsilon^3 h_3^{\rm B}(m_c) +O(\epsilon^4) \\
\int_{|m_c|/\epsilon}^\infinity d\xi \int_0^1 dt
\frac{ \check{k}^3t^2 \textrm{tanh}'\bb{{\epsilon E}/{2}}}{E^2(E^2\check{u}^2-t^2\xi^2\check{k}^2)} &=&  O(\epsilon^3) \label{intuBEC1}\\
\int_{|m_c|/\epsilon}^\infinity d\xi \int_0^1 dt
\frac{ \check{k}^3t^2 \textrm{tanh}'\bb{{\epsilon E}/{2}}\xi^2}{E^2(E^2\check{u}^2-t^2\xi^2\check{k}^2)} &=&  \epsilon\bbcro{-g_1^{\rm B} (m_c)+4\check{u}B(\check{u},m_c)} +O(\epsilon^2)
 \\
\int_{|\mu|/\Delta}^\infinity d\xi \int_0^1 dt
\frac{ \check{k}^3t^2 \textrm{tanh}'\bb{{\epsilon E}/{2}}\xi}{E^2(E^2\check{u}^2-t^2\xi^2\check{k}^2)} &=& \epsilon^2 C_2(\check{u},m_c) +O(\epsilon^3) \label{intuBEC3}
\eea
The integrals containing a resonant denominator are obtained after the decomposition $$\frac{\check{k}^3t^2 \textrm{tanh}'\bb{{\epsilon E}/{2}}}{E^2(E^2\check{u}^2-t^2\xi^2\check{k}^2)}=-\frac{\check{k}\textrm{tanh}'\bb{{\epsilon E}/{2}}}{\xi^2E^2}+\frac{\check{u}\check{k}\textrm{tanh}'\bb{{\epsilon E}/{2}}}{2\xi^2E}\bb{\frac{1}{E\check{u}-t\xi\check{k}}+\frac{1}{E\check{u}+t\xi\check{k}}}$$
Using this formulary, we obtain equations (\ref{mppTcBEC}--\ref{mpmTcBEC}) of the main text, with
\bea
\alpha_1(m_c)&=& |m_c|\bb{\frac{f_3^{\rm B}(m_c)}{6}-\frac{g_3^{\rm B}(m_c)}{12}}\\
\alpha_2(m_c)&=& -\frac{|m_c|\,f_1^{\rm B}(m_c)}{2} \\
\beta(m_c)&=& \frac{f_1^{\rm B}(m_c)}{2}- \frac{g_1^{\rm B}(m_c)}{4} \\
\gamma(m_c)&=& -\frac{h_1^{\rm B}(m_c)}{4} \\
C(\check{u},m_c) &=& -\frac{h_3^{\rm B}(m_c)}{4} + \frac{C_2(\check{u},m_c)}{8}
\eea

\section{Dynamic structure factor}
\label{structfactor}

The density-density response function is determined here within the Random
Phase Approximation, similarly to Refs. \cite{Castin2001,Zou2010}
and the BCS-Leggett response theory of Ref. \cite{He2016}.
The real-time density-density response is described by the retarded Green's function%
\begin{equation}
G_{\rho}^{R}\left(  \mathbf{q},\omega+i0^{+}\right)  \equiv-i\lim
_{\delta\rightarrow+0}\int_{0}^{\infty}e^{i\omega t-\delta t}\int
d\mathbf{r}~e^{-i\mathbf{q}\cdot\mathbf{r}}\left\langle \left[  \rho\left(
\mathbf{r},t\right)  ,\rho\left(  0,0\right)  \right]  \right\rangle dt,
\label{GR}%
\end{equation}
where $\rho\left(  \mathbf{r},t\right)  $ is the particle density:%
\begin{equation}
\rho\left(  \mathbf{r},t\right)  =\bar{\psi}_{\mathbf{r},t,\uparrow}%
\psi_{\mathbf{r},t,\uparrow}+\bar{\psi}_{\mathbf{r},t,\downarrow}%
\psi_{\mathbf{r},t,\downarrow}. \label{dens}%
\end{equation}
The spectral weight function (the dynamic structure factor) is proportional to
the imaginary part of $G_{\rho}^{R}$:%
\begin{equation}
\chi_{\rho}\left(  \mathbf{q},\omega\right)  =-\frac{1}{\pi}\operatorname{Im}%
G_{\rho}^{R}\left(  \mathbf{q},\omega+i0^{+}\right)  . \label{Srho}%
\end{equation}
We use the known correspondence between the Green's function in the Matsubara
representation $\mathcal{G}_{\rho}\left(  \mathbf{q},i\Omega_{m}\right)  $ and
the retarded two-point Green's function $G_{\rho}^{R}\left(  \mathbf{q}%
,\omega+i0^{+}\right)  $ [e.~g., \cite{Mahan}, Eq. (3.3.11)]:%
\begin{equation}
G_{\rho}^{R}\left(  \mathbf{q},\omega+i0^{+}\right)  =~\underset{i\Omega
_{m}\rightarrow\omega+i0^{+}}{\mathcal{G}_{\rho}\left(  \mathbf{q},i\Omega
_{m}\right)  .} \label{corresp}%
\end{equation}

The Green's function in the Matsubara representation $\mathcal{G}_{\rho
}\left(  \mathbf{q},i\Omega_{m}\right)  $ is determined using the generating
functional in the path-integral representation with the auxiliary
infinitesimal field variable $\upsilon\left(  \mathbf{r},\tau\right)  $
corresponding to density fluctuations,%
\[
\Xi\left(  \upsilon\right)  =\left\langle \exp\left[  \int_{0}^{\beta}%
d\tau\int d\mathbf{r}~\upsilon\left(  \mathbf{r},\tau\right)  \rho\left(
\mathbf{r},\tau\right)  \right]  \right\rangle _{S}%
\]
where the action functional $S$ is given by Eq. (\ref{S}). The next derivation
is standard, as described in the main text: (1) introducing the pair field
$\left[  \bar{\Psi},\Psi\right]  $, (2) performing the Hubbard-Stratonovich
transformation, (3) integrating out the fermion fields, (4) expanding the
effective bosonic action up to quadratic order in the pair field and density
fluctuations. This leads to the generating functional as the path-integral
average with the \emph{bosonic} GPF action:%
\begin{equation}
\Xi\left[  \upsilon\right]  =\left\langle \exp\left(  -S_{\upsilon}\left[
\upsilon\right]  \right)  \right\rangle _{S_{GPF}},\label{Xi}%
\end{equation}
where the auxiliary action $\tilde{S}_{\upsilon}\left[  \upsilon\right]  $ is
useful to be written in the modulus-phase basis for fluctuation variables
$\varphi_{\mathbf{q},n}$,%
\begin{equation}
\lambda_{\mathbf{q},m}=\frac{\phi_{\mathbf{q},m}+\bar{\phi}_{-\mathbf{q},-m}%
}{\sqrt{2}},\qquad\theta_{\mathbf{q},m}=\frac{\phi_{\mathbf{q},m}-\bar{\phi
}_{-\mathbf{q},-m}}{\sqrt{2}i}.\label{phas}%
\end{equation}
The resulting auxiliary action is:%
\begin{align}
S_{\upsilon}\left[  \upsilon\right]   &  =\frac{1}{2}\sum_{\mathbf{q}%
,m}\left\{  {M}_{\rho\rho}\left(  \mathbf{q},i\Omega_{m}\right)
\upsilon_{\mathbf{q},m}\upsilon_{-\mathbf{q},-m}\right.  \nonumber\\
&  \left.  +2\left[  {M}_{-\rho}\left(  \mathbf{q},i\Omega_{m}\right)
\bar{\lambda}_{\mathbf{q},m}-{M}_{+\rho}\left(  \mathbf{q},i\Omega
_{m}\right)  i \bar{\theta}_{\mathbf{q},m}\right]  \upsilon_{\mathbf{q}%
,m}\right\}  ,
\end{align}
with the matrix elements (we use the convention $2m=k_F=\epsilon_F=1$ everywhere in this appendix):%
\begin{align}
{M}_{-\rho}\left(  \mathbf{q},i\Omega_{m}\right)   &  =\sqrt{2}\Delta
\int\frac{d\mathbf{k}}{\left(  2\pi\right)  ^{3}}\frac{X\left(  E_{\mathbf{k}%
}\right)  }{4E_{\mathbf{k}}E_{\mathbf{k}+\mathbf{q}}}\left(  \xi_{\mathbf{k}%
}+\xi_{\mathbf{k}+\mathbf{q}}\right)  \nonumber\\
&  \times\left(  \frac{1}{i\Omega_{m}-E_{\mathbf{k}}-E_{\mathbf{k}+\mathbf{q}%
}}-\frac{1}{i\Omega_{m}+E_{\mathbf{k}}+E_{\mathbf{k}+\mathbf{q}}}\right.
\nonumber\\
&  \left.  +\frac{1}{i\Omega_{m}+E_{\mathbf{k}}-E_{\mathbf{k}+\mathbf{q}}%
}-\frac{1}{i\Omega_{m}-E_{\mathbf{k}}+E_{\mathbf{k}+\mathbf{q}}}\right)
,\label{Q13}%
\end{align}

\begin{align}
{M}_{+\rho}\left(  \mathbf{q},i\Omega_{m}\right)   &  ={\sqrt
{2}\Delta}\int\frac{d\mathbf{k}}{\left(  2\pi\right)  ^{3}}\frac{X\left(
E_{\mathbf{k}}\right)  }{4E_{\mathbf{k}}E_{\mathbf{k}+\mathbf{q}}}\nonumber\\
&  \times\left[  \left(  E_{\mathbf{k}+\mathbf{q}}+E_{\mathbf{k}}\right)
\left(  \frac{1}{i\Omega_{m}-E_{\mathbf{k}}-E_{\mathbf{k}+\mathbf{q}}}%
+\frac{1}{i\Omega_{m}+E_{\mathbf{k}}+E_{\mathbf{k}+\mathbf{q}}}\right)
\right. \nonumber\\
&  \left.  +\left(  E_{\mathbf{k}+\mathbf{q}}-E_{\mathbf{k}}\right)  \left(
\frac{1}{i\Omega_{m}+E_{\mathbf{k}}-E_{\mathbf{k}+\mathbf{q}}}+\frac
{1}{i\Omega_{m}-E_{\mathbf{k}}+E_{\mathbf{k}+\mathbf{q}}}\right)  \right]  ,
\label{Q23}%
\end{align}
and%
\begin{align}
{M}_{\rho\rho}\left(  \mathbf{q},i\Omega_{m}\right)   &  =\int\frac
{d\mathbf{k}}{\left(  2\pi\right)  ^{3}}\frac{X\left(  E_{\mathbf{k}}\right)
}{2E_{\mathbf{k}}E_{\mathbf{k}+\mathbf{q}}}\nonumber\\
&  \times\left(  \frac{E_{\mathbf{k}}E_{\mathbf{k}+\mathbf{q}}-\xi
_{\mathbf{k}}\xi_{\mathbf{k}+\mathbf{q}}+\Delta^{2}}{i\Omega_{m}%
-E_{\mathbf{k}}-E_{\mathbf{k}+\mathbf{q}}}+\frac{E_{\mathbf{k}}E_{\mathbf{k}%
+\mathbf{q}}+\xi_{\mathbf{k}}\xi_{\mathbf{k}+\mathbf{q}}-\Delta^{2}}%
{i\Omega_{m}-E_{\mathbf{k}}+E_{\mathbf{k}+\mathbf{q}}}\right. \nonumber\\
&  \left.  -\frac{E_{\mathbf{k}}E_{\mathbf{k}+\mathbf{q}}+\xi_{\mathbf{k}}%
\xi_{\mathbf{k}+\mathbf{q}}-\Delta^{2}}{i\Omega_{m}+E_{\mathbf{k}%
}-E_{\mathbf{k}+\mathbf{q}}}-\frac{E_{\mathbf{k}}E_{\mathbf{k}+\mathbf{q}}%
-\xi_{\mathbf{k}}\xi_{\mathbf{k}+\mathbf{q}}+\Delta^{2}}{i\Omega
_{m}+E_{\mathbf{k}}+E_{\mathbf{k}+\mathbf{q}}}\right)  . \label{Q33}%
\end{align}

The retarded Green's function in the Matsubara representation is determined
by:%
\begin{equation}
\mathcal{G}_{\rho}\left(  \mathbf{q},i\Omega_{m}\right)  =-\left.
\frac{\partial^{2}\Xi\left[  \upsilon\right]  }{\partial\bar{\upsilon
}_{\mathbf{q},m}\partial\upsilon_{\mathbf{q},m}}\right\vert _{\upsilon=0},
\end{equation}
which results in expression \eqref{Gr} given in the main text.

The long-wavelength expansion of matrix elements (\ref{Q13}) to (\ref{Q33})
for an arbitrary complex $u$ [used in (\ref{lwlim}) with $u=c+i0^{+}$] gives
us the results:%
\begin{align}
\lim_{q\rightarrow0}{M}_{-\rho}\left(  \mathbf{q},uq\right)   &
=-\frac{\sqrt{2}\Delta}{4\pi^{2}}\int_{0}^{\infty}k^{2}dk\frac{\xi
_{\mathbf{k}}}{E_{\mathbf{k}}^{2}}\left\{  \frac{X\left(  E_{\mathbf{k}%
}\right)  }{E_{k}}\right.  \nonumber\\
&  +\left.  X^{\prime}\left(  E_{\mathbf{k}}\right)  \left[  \frac{E_{k}%
u}{2k\xi_{k}}\operatorname{arctanh}\left(  \frac{2k\xi_{k}}{E_{k}u}\right)
-1\right]  \right\}  ,
\end{align}%
\begin{align}
\lim_{q\rightarrow0}{M}_{+\rho}\left(  \mathbf{q},uq\right)   &
=-\frac{\sqrt{2}\Delta uq}{8\pi^{2}}\int_{0}^{\infty}k^{2}dk~\frac
{1}{E_{\mathbf{k}}^{2}}\left\{  \frac{X\left(  E_{\mathbf{k}}\right)  }{E_{k}%
}\right.  \nonumber\\
&  +\left.  X^{\prime}\left(  E_{\mathbf{k}}\right)  \left[  \frac{E_{k}%
u}{2k\xi_{k}}\operatorname{arctanh}\left(  \frac{2k\xi_{k}}{E_{k}u}\right)
-1\right]  \right\}  ,
\end{align}%
\begin{align}
\lim_{q\rightarrow0}{M}_{\rho\rho}\left(  \mathbf{q},uq\right)   &
=-\frac{1}{2\pi^{2}}\int_{0}^{\infty}k^{2}dk\frac{1}{E_{\mathbf{k}}^{2}%
}\left\{  \Delta^{2}\frac{X\left(  E_{\mathbf{k}}\right)  }{E_{k}}\right.
\nonumber\\
&  -\left.  \xi_{\mathbf{k}}^{2}X^{\prime}\left(  E_{\mathbf{k}}\right)
\left[  \frac{E_{k}u}{2k\xi_{k}}\operatorname{arctanh}\left(  \frac{2k\xi_{k}%
}{E_{k}u}\right)  -1\right]  \right\}  .
\end{align}

There is in fact an analytic expression of ``pure density'' contribution  $\chi_{\rho}^{\left(
1\right)  }\left(  c\right)  $ to the density response:%
\begin{equation}
\chi_{\rho}^{\left(  1\right)  }\left(  c\right)  =\frac{1}{16\pi^{2}%
}c\left\{
\begin{array}
[c]{cc}%
1-X\left(  E_{k_{3}}\right)  -X\left(  E_{k_{2}}\right)  +X\left(  E_{k_{1}%
}\right)  , & c<c_{b}\\
1-X\left(  E_{k_{3}}\right)  , & c\geq c_{b}%
\end{array}
\right.  \label{hi1}%
\end{equation}
where boundary values for the momentum $k_{1},k_{2},k_{3}$ and the boundary
velocity $c_{b}$ are described in Sec. \ref{sec:analcont}.




\end{document}